\documentclass[epj]{svjour}
\usepackage{graphics}
\onecolumn

\begin{document}
\title{The ``Closed'' Chiral Symmetry and Its Application to Tetraquark}

\author{Hua-Xing Chen
}                     
\institute{School of Physics and Nuclear Energy Engineering, Beihang University, Beijing 100191, China}
\date{Received: date / Revised version: date}
%
\abstract{We investigate the chiral (flavor) structure of tetraquarks, and study chiral transformation properties of the ``non-exotic'' $[(\bar \mathbf{3}, \mathbf{3})\oplus(\mathbf{3}, \bar \mathbf{3})]$ and $[({\mathbf 8},{\mathbf 1}) \oplus ({\mathbf 1},{\mathbf 8})]$ tetraquark chiral multiplets.
We find that as long as this kind of tetraquark states contains one quark and one antiquark having the same chirality, such as $q_L q_L \bar q_L \bar q_R + q_R q_R \bar q_R \bar q_L$, they transform in the same way as the lowest level $\bar q q$ chiral multiplets under chiral transformations. There is only one $[(\bar \mathbf{3}, \mathbf{3})\oplus(\mathbf{3}, \bar \mathbf{3})]$ chiral multiplet whose quark-antiquark pairs all have the opposite chirality ($q_L q_L \bar q_R \bar q_R + q_R q_R \bar q_L \bar q_L$), and it transforms differently from others. Based on these studies, we construct local tetraquark currents belonging to the ``non-exotic'' chiral multiplet $[(\bar \mathbf{3}, \mathbf{3})\oplus(\mathbf{3}, \bar \mathbf{3})]$ and having quantum numbers $J^{PC}=1^{-+}$.
} 
\maketitle

%
\section{Introduction}
\label{intro}

The spontaneously breaking of chiral symmetry is one of the key issues to understand the nonlinear behaviors of the Quantum Chromodynamics (QCD) at the low energy region as well as the mass of hadrons~\cite{Hatsuda:1994pi,lee72,lee81,Beane:2002td,Benmerrouche:1989uc,Haberzettl:1998rw,Cohen:2002st,DeTar:1988kn,Diakonov:1987ty,Weinberg:1969hw,Nagata:2007di,Jido:1999hd,Jido:2001nt}. To describe the chiral symmetry we use the $SU(3)_L \otimes SU(3)_R$ chiral group, which spontaneously breaks to the $SU(3)_V$ group and produces eight Goldstone bosons, the pseudoscalar mesons $\pi$, K and $\eta_8$. To study this chiral group we can use an algebraic method, and perform chiral transformations to study the chiral structure of hadrons~\cite{Weinberg:1969hw,Nagata:2007di,Jido:1999hd,Jido:2001nt,Cohen:1996sb,Chen:2008qv,Leinweber:1994nm}. During this process the chiral representation itself is not changed. For example, we can perform chiral transformations on baryons belonging to the $[(\mathbf 6, \mathbf 3)\oplus(\mathbf 3, \mathbf 6)]$ chiral representation, and the transformed ones still belong to the same representation. So the chiral symmetry is ``closed'' under chiral transformations.

The study of exotic hadrons is another interesting subject~\cite{Jaffe:1976ig,Weinstein:1982gc,Close:2002zu,Lipkin:1986dw,Aerts:1979hn,Brodsky:1977bs,Ping:2009zzb,Lee:2009rt,Vijande:2007ix,Oka:2004xh,Zhu:2007wz,MartinezTorres:2008gy}. Hadrons having exotic structures are called exotica, such as hybrid states, glueballs, tetraquark states and molecular states, etc.. The existence of exotic hadrons is not forbidden by QCD. However, there are quite a few experiments which have observed exotic hadrons. These observed exotica are $\pi_1(1400)$, $\pi_1(1600)$ and $\pi_1(2015)$, etc.~\cite{experiments}, but still they do not have an exotic flavor structure. Hence, the exotic flavor structure seems missing in our world. The chiral symmetry has been studied by lots of physicists, but the chiral structure of exotic hadrons has not been investigated by so many physicists yet. Since there are always multi-quark components in the Fock space expansion of physical hadron states, it is worth studying what remains unchanged in this expansion.

Here we would like to propose that the ``closed'' chiral symmetry may be related to the missing of the exotic flavor structure. Take the lowest level chiral multiplet ($\sigma$, $\pi$) as an example. Here we assume that they belong to the $[({\mathbf 3}, \bar {\mathbf 3}) + (\bar {\mathbf 3}, {\mathbf 3})]$ chiral representation. Their behaviors under the $U(1)_V$, $U(1)_A$, $SU(3)_V$ and $SU(3)_A$ chiral transformations reflect their scattering properties with the revelent pseudoscalar mesons. Since the transformed fields still belong to the same representation, the final states of this scattering should also belong to this chiral multiplet. They can be higher lever chiral multiplets, such as $(\rho, a_1)$, but they still have a non-exotic flavor structure.

Take the meson $f_1$ of $J^{PC} = 1^{++}$ as another example. We use the interpolating field $J_{\mu} = \bar q \gamma_\mu \gamma_5 q$ as the lowest level term, which can couple to $f_1$. It belongs to the chiral representation $[({\mathbf 8}, {\mathbf 1}) + ({\mathbf 1}, {\mathbf 8})]$. Its full Fock space expansion can be written as:
\begin{eqnarray}\label{eq:fock}
| f_1 \rangle &=& | \bar q \gamma_\mu \gamma_5 q \rangle + | \bar q \gamma_\nu G^{\mu\nu} q \rangle + | \bar q \gamma_\mu \gamma_5 q \bar q q \rangle + \cdots \, ,
\end{eqnarray}
where latter terms can be obtained {\it by properly adding a chiral singlet quark-antiquark pair or a (chiral singlet) gluon to some former terms}. All these terms belong to the same $[({\mathbf 8}, {\mathbf 1}) + ({\mathbf 1}, {\mathbf 8})]$ chiral multiplet. Chiral transformations do not change this chiral representation, and so there are no exotic flavor structures in this expansion. However, the third term $\bar q \gamma_\mu \gamma_5 q \bar q q$ can be transformed to $\bar q \gamma_\mu \gamma_5 q \bar q \gamma_5 q$, which can have exotic quantum numbers $J^{PC} = 1^{-+}$ and so can couple to the exotic meson $\pi_1$.
We note that the exotic flavor structure may still exist, but then it can not be transformed to, nor transformed from, any term contained in the Fock space expansion of ground mesons and baryons under chiral transformations, i.e., it is not the chiral partner of any observed physical hadron. This may make them difficult to be observed.

In this paper we shall study the chiral structure of the tetraquark state. We shall only consider the chiral (flavor) degree of freedom and leave others undetermined, such as the color, orbit and spin. These degrees of freedom can be fixed in other models under proper assumptions. We shall perform chiral transformations on the tetraquarks belonging to the ``non-exotic'' $[({\mathbf 3}, \bar {\mathbf 3}) + (\bar {\mathbf 3}, {\mathbf 3})]$ and the $[({\mathbf 8}, {\mathbf 1}) + ({\mathbf 1}, {\mathbf 8})]$ chiral representations. Some of them contain one pair of quark and antiquark which have the same chirality and are combined to be a chiral singlet ($\bar q^A_L q^A_L + \bar q^A_R q^A_R$). These tetraquarks transform in the same way exactly as the lowest level $\bar q q$ mesons belonging to the $[({\mathbf 3}, \bar {\mathbf 3}) + (\bar {\mathbf 3}, {\mathbf 3})]$ or the $[({\mathbf 8}, {\mathbf 1}) + ({\mathbf 1}, {\mathbf 8})]$ chiral multiplets, or their mirror multiplets. This seems to be trivial but the flavor indices need to be organized in a proper way in order to construct a tetraquark state belonging to these representations. Except these $[({\mathbf 3}, \bar {\mathbf 3}) + (\bar {\mathbf 3}, {\mathbf 3})]$ and $[({\mathbf 8}, {\mathbf 1}) + ({\mathbf 1}, {\mathbf 8})]$ chiral multiplets, there is only one $[(\bar {\mathbf 3}, {\mathbf 3}) + ({\mathbf 3}, \bar {\mathbf 3})]$ chiral multiplet, whose quark-antiquark pairs all have the opposite chirality ($q_L q_L \bar q_R \bar q_R + q_R q_R \bar q_L \bar q_L$), and it transforms differently from other multiplets.


This paper is organized as follows. In Sec.~\ref{sec:flavor} we study the flavor structure of tetraquarks. In Sec.~\ref{sec:chiral} we study the chiral structure of tetraquarks. We separately investigate tetraquarks belonging to the flavor singlet, flavor decuplet, flavor anti-decuplet, flavor $\mathbf{27}$ and flavor octet. In Sec.~\ref{sec:transformation} we study chiral transformations of the $[({\mathbf 3},\bar{\mathbf 3}) \oplus (\bar{\mathbf 3},{\mathbf 3})]$ and $[({\mathbf 8},{\mathbf 1}) \oplus ({\mathbf 1},{\mathbf 8})]$ chiral multiplets. Based on these studies, in Sec.~\ref{sec:current} we construct local tetraquark currents belonging to the ``non-exotic'' chiral multiplet $[(\bar \mathbf{3}, \mathbf{3})\oplus(\mathbf{3}, \bar \mathbf{3})]$ and having quantum numbers $J^{PC}=1^{-+}$. Sec.~\ref{sec:summary} is a summary.

\section{Flavor Structure of Tetraquark}
\label{sec:flavor}

\begin{figure}[hbt]
\begin{center}
\scalebox{0.4}{\includegraphics{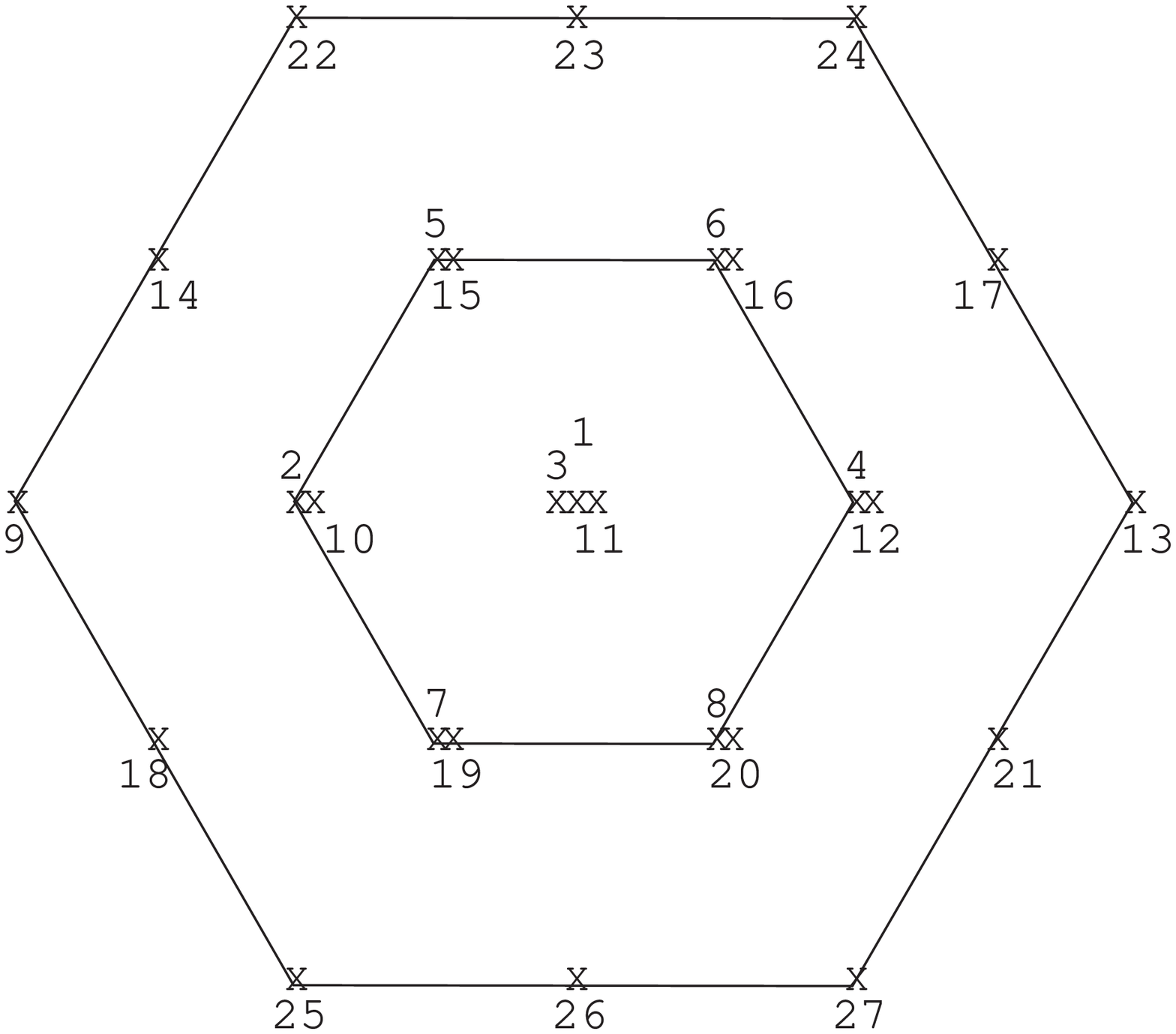}}
\caption{Weight diagram of the $\mathbf{27}$ flavor representation.} \label{fig:27weight}
\end{center}
\end{figure}

The flavor structure of tetraquark is:
\begin{eqnarray}\label{eq:flavor}
\nonumber \mathbf{3}\otimes\mathbf{3} \otimes \mathbf{\bar 3}\otimes\mathbf{\bar3}
&=& \Big( \mathbf{\bar 3}\oplus\mathbf{6} \Big) \otimes \Big( \mathbf{3}\oplus\mathbf{\bar6} \Big)
\\ = \Big( \mathbf{\bar 3}\otimes\mathbf{3} \Big) &\oplus& \Big( \mathbf{\bar 3}\otimes\mathbf{\bar 6} \Big)
\oplus \Big( \mathbf{6}\otimes\mathbf{3} \Big) \oplus \Big( \mathbf{6}\otimes\mathbf{\bar 6} \Big)
\\ \nonumber = \Big( \mathbf{1}\oplus \mathbf{8}\Big) &\oplus& \Big( \mathbf{8}\oplus\mathbf{\overline{10}} \Big)
\oplus \Big( \mathbf{8}\oplus \mathbf{10}\Big) \oplus \Big( \mathbf{1}\oplus \mathbf{8}\oplus\mathbf{27} \Big) \, .
\end{eqnarray}
In our following calculation we shall use $A,B,C,\cdots$ to denote flavor indices. $\epsilon^{ABC}$ is the totally anti-symmetric tensor; $S_P^{ABC}$ ($P=1\cdots10$) are the normalized totally symmetric matrices; $\bf \lambda_N$ ($N=1\cdots8$) are the Gell-Mann matrices; $S^{ABCD}_{U}$ ($U=1\cdots27$) are the matrices for the $\mathbf{27}$ flavor representation. We show their non-zero components in Tab~\ref{tab:S27}. The weight diagram of this representation is a hexagon with three layers as shown in Fig.~\ref{fig:27weight}.
In our definitions $S^{ABCD}_{1}$ denotes the only one central point which is the inner layer, $S^{ABCD}_{2\cdots8}$ denote the six points in the intermediate layer together with another point in the inner layer, and $S^{ABCD}_{9\cdots27}$ denote others twelve points in the outer layer together with other six points in the intermediate layer and one point in the inner layer. They satisfy the normalization conditions:
\begin{eqnarray}
\sum_A \sum_B \sum_C \sum_D |S^{ABCD}_{U}|^2 = 1 \, ,
\end{eqnarray}
and the following commutation and orthogonality relations:
\begin{eqnarray}
&& S^{ABCD}_{U} = S^{BACD}_{U} = S^{ABDC}_{U} \, ,
\\ \nonumber && \sum_A \sum_B \sum_C \sum_D |S^{ABCD}_{U}S^{ABCD}_{V}| = 0 \, , \mbox{ for } \, U \neq V \, .
\end{eqnarray}

\begin{table}[tbh]
\renewcommand{\arraystretch}{1.3}
\begin{center}
\caption{Non-zero components of $S^{ABCD}_{U}$.}
\begin{tabular}{|c||p{2cm}|c|p{2cm}|c|p{2cm}|c|p{2cm}|c|}
\hline \hline $U$ & \hfil{$ABCD$} & $S^{ABCD}_{U}$ & \hfil{$ABCD$} & $S^{ABCD}_{U}$ & \hfil{$ABCD$} & $S^{ABCD}_{U}$ & \hfil{$ABCD$} & $S^{ABCD}_{U}$
\\ \hline \hline 1 & \hfil{1111, 2222} & $1/\sqrt{30}$ & \hfil{3333} & $\sqrt{3/10}$ &
\hfil{1212} & $1/\sqrt{120}$ & \hfil{1313,2323} & $-\sqrt{3/40}$
\\ \hline \hline 2 & \multicolumn{3}{c|}{1112,1222} & $1/\sqrt{20}$ & \multicolumn{3}{c|}{1323} & $-1/\sqrt{5}$
\\ \hline 3 & \multicolumn{3}{c|}{1111,2323} & $1/\sqrt{10}$ & \multicolumn{3}{c|}{2222,1313} & $-1/\sqrt{10}$
\\ \hline 4 & \multicolumn{3}{c|}{1211,2212} & $1/\sqrt{20}$ & \multicolumn{3}{c|}{2313} & $-1/\sqrt{5}$
\\ \hline 5 & \hfil{1113} & $\sqrt{2/15}$ & \hfil{1333} & $-\sqrt{3/10}$ & \multicolumn{3}{c|}{1223} & $1/\sqrt{30}$
\\ \hline 6 & \hfil{2223} & $\sqrt{2/15}$ & \hfil{2333} & $-\sqrt{3/10}$ & \multicolumn{3}{c|}{1213} & $1/\sqrt{30}$
\\ \hline 7 & \hfil{2322} & $\sqrt{2/15}$ & \hfil{3323} & $-\sqrt{3/10}$ & \multicolumn{3}{c|}{1312} & $1/\sqrt{30}$
\\ \hline 8 & \hfil{1311} & $\sqrt{2/15}$ & \hfil{3313} & $-\sqrt{3/10}$ & \multicolumn{3}{c|}{2312} & $1/\sqrt{30}$
\\ \hline \hline 9 & \multicolumn{3}{c|}{1122} & $1$ & \multicolumn{4}{c|}{}
\\ \hline 10 & \multicolumn{3}{c|}{1112} & $1/2$ & \multicolumn{3}{c|}{1222} & $-1/2$
\\ \hline 11 & \multicolumn{3}{c|}{1111,2222} & $1/\sqrt6$ & \multicolumn{3}{c|}{1212} & $-1/\sqrt6$
\\ \hline 12 & \multicolumn{3}{c|}{1211} & $1/2$ & \multicolumn{3}{c|}{2212} & $-1/2$
\\ \hline 13 & \multicolumn{3}{c|}{2211} & $1$ & \multicolumn{4}{c|}{}
\\ \hline 14 & \multicolumn{3}{c|}{1123} & $1/\sqrt2$ & \multicolumn{4}{c|}{}
\\ \hline 15 & \multicolumn{3}{c|}{1113} & $1/\sqrt6$ & \multicolumn{3}{c|}{1223} & $-1/\sqrt6$
\\ \hline 16 & \multicolumn{3}{c|}{2223} & $-1/\sqrt6$ & \multicolumn{3}{c|}{1213} & $1/\sqrt6$
\\ \hline 17 & \multicolumn{3}{c|}{2213} & $1/\sqrt2$ & \multicolumn{4}{c|}{}
\\ \hline 18 & \multicolumn{3}{c|}{1322} & $1/\sqrt2$ & \multicolumn{4}{c|}{}
\\ \hline 19 & \multicolumn{3}{c|}{2322} & $-1/\sqrt6$ & \multicolumn{3}{c|}{1312} & $1/\sqrt6$
\\ \hline 20 & \multicolumn{3}{c|}{1311} & $1/\sqrt6$ & \multicolumn{3}{c|}{2312} & $-1/\sqrt6$
\\ \hline 21 & \multicolumn{3}{c|}{2311} & $1/\sqrt2$ & \multicolumn{4}{c|}{}
\\ \hline 22 & \multicolumn{3}{c|}{1133} & $1$  & \multicolumn{4}{c|}{}
\\ \hline 23 & \multicolumn{3}{c|}{1233} & $1/\sqrt2$ & \multicolumn{4}{c|}{}
\\ \hline 24 & \multicolumn{3}{c|}{2233} & $1$ & \multicolumn{4}{c|}{}
\\ \hline 25 & \multicolumn{3}{c|}{3322} & $1$ & \multicolumn{4}{c|}{}
\\ \hline 26 & \multicolumn{3}{c|}{3312} & $1/\sqrt2$ & \multicolumn{4}{c|}{}
\\ \hline 27 & \multicolumn{3}{c|}{3311} & $1$ & \multicolumn{4}{c|}{}
\\ \hline \hline
\end{tabular}
\label{tab:S27}
\end{center}
\renewcommand{\arraystretch}{1.0}
\end{table}

We can explicitly write down all the tetraquark flavor representations shown in Eq.~(\ref{eq:flavor}) using these definitions and notations. We do this according to the flavor symmetries of diquarks and antidiquarks:
\begin{enumerate}

\item When the diquark and the antidiquark both have the antisymmetric flavor structure, i.e., the flavor representation of $(qq)(\bar q \bar q)$ is $\mathbf{\bar 3} \otimes \mathbf{3}$, the flavor representations of tetraquark can be singlet ($\mathcal{S}_1$) and octet ($\mathcal{O}_1^N$):
\begin{eqnarray}
\mathcal{S}_1 &=& \epsilon^{ABE}\epsilon^{CDE} (q_A q_B)(\bar q_C \bar q_D)
\\ \nonumber &=& (q_A q_B)(\bar q_A \bar q_B - \bar q_B \bar q_A) \, ,
\\ \nonumber \mathcal{O}_1^N &=& \big ( \delta^{AC} \lambda_N^{DB} - \delta^{BC} \lambda_N^{DA} - \delta^{AD} \lambda_N^{CB} + \delta^{BD} \lambda_N^{CA} \big ) \times
\\ \nonumber && ~~~~~~~~~~~~~~~~~~~~~~~~~~~~~~~~~~~~~~ \times (q_A q_B)(\bar q_C \bar q_D)
\\ \nonumber &=& \delta^{AC} \lambda_N^{DB} (q_A q_B - q_B q_A)(\bar q_C \bar q_D - \bar q_D \bar q_C) \, .
\end{eqnarray}

\item When the diquark has the antisymmetric flavor structure and the antidiquark has the symmetric structure ($\mathbf{\bar 3} (qq) \otimes \mathbf{\bar 6} (\bar q \bar q)$), the flavor representations of tetraquark can be octet ($\mathcal{O}_2^N$) and anti-decuplet ($\overline{\mathcal{D}}^P$):
\begin{eqnarray}
\nonumber \mathcal{O}_2^N &=& \big ( \delta^{AC} \lambda_N^{DB} - \delta^{BC} \lambda_N^{DA} + \delta^{AD} \lambda_N^{CB} - \delta^{BD} \lambda_N^{CA} \big ) \times
\\ \nonumber && ~~~~~~~~~~~~~~~~~~~~~~~~~~~~~~~~~~~~~~ \times (q_A q_B)(\bar q_C \bar q_D)
\\ &=& \delta^{AC} \lambda_N^{DB} (q_A q_B - q_B q_A)(\bar q_C \bar q_D + \bar q_D \bar q_C) \, ,
\\ \nonumber \overline{\mathcal{D}}^P &=& \epsilon^{ABE}S_P^{CDE} (q_A q_B)(\bar q_C \bar q_D) \, .
\end{eqnarray}

\item When the diquark has the symmetric flavor structure and the antidiquark has the antisymmetric structure ($\mathbf{6} (qq) \otimes \mathbf{3} (\bar q \bar q)$), the flavor representations of tetraquark can be octet ($\mathcal{O}_3^N$) and decuplet ($\mathcal{D}^P$):
\begin{eqnarray}
\nonumber \mathcal{O}_3^N &=& \big ( \delta^{AC} \lambda_N^{DB} + \delta^{BC} \lambda_N^{DA} - \delta^{AD} \lambda_N^{CB} - \delta^{BD} \lambda_N^{CA} \big ) \times
\\ \nonumber && ~~~~~~~~~~~~~~~~~~~~~~~~~~~~~~~~~~~~~~ \times (q_A q_B)(\bar q_C \bar q_D)
\\ &=& \delta^{AC} \lambda_N^{DB} (q_A q_B + q_B q_A)(\bar q_C \bar q_D - \bar q_D \bar q_C) \, ,
\\ \nonumber \mathcal{D}^P &=& S_P^{ABE} \epsilon^{CDE} (q_A q_B)(\bar q_C \bar q_D) \, .
\end{eqnarray}

\item When the diquark and the antidiquark both have the symmetric flavor structure ($\mathbf{6} (qq) \otimes \mathbf{\bar6} (\bar q \bar q)$), the flavor representations of tetraquark can be singlet ($\mathcal{S}_2$), octet ($\mathcal{O}_4^N$) and $\mathbf{27}_f$ ($\mathcal{T}^U$):
\begin{eqnarray}
\nonumber \mathcal{S}_2 &=& ( \delta^{AC} \delta^{BD} + \delta^{AD} \delta^{BC} )(q_A q_B)(\bar q_C \bar q_D)
\\ &=& (q_A q_B)(\bar q_A \bar q_B+\bar q_B \bar q_A) \, ,
\\ \nonumber \mathcal{O}_4^N &=& \big ( \delta^{AC} \lambda_N^{DB} + \delta^{BC} \lambda_N^{DA} + \delta^{AD} \lambda_N^{CB} + \delta^{BD} \lambda_N^{CA} \big ) \times
\\ \nonumber && ~~~~~~~~~~~~~~~~~~~~~~~~~~~~~~~~~~~~~~ \times (q_A q_B)(\bar q_C \bar q_D)
\\ \nonumber &=& \delta^{AC} \lambda_N^{DB} (q_A q_B + q_B q_A)(\bar q_C \bar q_D + \bar q_D \bar q_C) \, ,
\\ \nonumber \mathcal{T}^U &=& S_U^{ABCD} (q_A q_B)(\bar q_C \bar q_D) \, .
\end{eqnarray}

\end{enumerate}

\section{Chiral Structure of Tetraquark}
\label{sec:chiral}

Compared with its flavor structure, the chiral structure of the tetraquark is more complicated:
\begin{eqnarray}
&& \Big( ({\mathbf 3},{\mathbf 1})\oplus({\mathbf 1},{\mathbf 3}) \Big)^2 \otimes \Big( (\bar{\mathbf 3},{\mathbf 1})\oplus({\mathbf 1},\bar{\mathbf 3}) \Big)^2
\\ \nonumber &=& \Big( (\bar{\mathbf 3},{\mathbf 1})\oplus({\mathbf 1},\bar{\mathbf 3}) \oplus ({\mathbf 6},{\mathbf 1})\oplus({\mathbf 1},{\mathbf 6}) \oplus ({\mathbf 3},{\mathbf 3}) \oplus ({\mathbf 3},{\mathbf 3}) \Big)
\otimes \Big( ({\mathbf 3},{\mathbf 1})\oplus({\mathbf 1},{\mathbf 3}) \oplus (\bar{\mathbf 6},{\mathbf 1})\oplus({\mathbf 1},\bar{\mathbf 6}) \oplus (\bar{\mathbf 3},\bar{\mathbf 3}) \oplus (\bar{\mathbf 3},\bar{\mathbf 3}) \Big)
\end{eqnarray}
We show its full expression in Tab~\ref{tab:chiral}, where $q_L$ and $q_R$ denote the left-handed and right-handed quarks, respectively. We note that the chiral structure of the tetraquark is always like $[({\mathbf A},{\mathbf B})+({\mathbf B},{\mathbf A})]$, since ``left'' and ``right'' can be interchanged in the QCD sector.

\begin{table}[!hbt]
\renewcommand{\arraystretch}{1.2}
\begin{center}
\caption{The chiral structure of tetraquarks.}
\begin{tabular}{c c c c}
\hline\hline
Tetraquark Chirality & Chiral Multiplets & Diquark Symmetry ($q q \otimes \bar q \bar q$)
\\ \hline \hline
$q_L q_L \bar q_L \bar q_L + q_R q_R \bar q_R \bar q_R$
& $\begin{array}{c}
[({\mathbf 1},{\mathbf 1}) \oplus ({\mathbf 1},{\mathbf 1})] \, , \, [({\mathbf 8},{\mathbf 1}) \oplus ({\mathbf 1},{\mathbf 8})]
\\ \hline [({\mathbf 8},{\mathbf 1}) \oplus ({\mathbf 1},{\mathbf 8})] \, , \, [(\overline{\mathbf {10}},{\mathbf 1}) \oplus ({\mathbf 1},\overline{\mathbf {10}})]
\\ \hline [({\mathbf 8},{\mathbf 1}) \oplus ({\mathbf 1},{\mathbf 8})] \, , \, [(\mathbf {10},{\mathbf 1}) \oplus ({\mathbf 1},\mathbf {10})]
\\ \hline [({\mathbf 1},{\mathbf 1}) \oplus ({\mathbf 1},{\mathbf 1})] , [({\mathbf 8},{\mathbf 1}) \oplus ({\mathbf 1},{\mathbf 8})] , [(\mathbf{27},{\mathbf 1}) \oplus ({\mathbf 1},\mathbf{27})]
\end{array}$
& $\begin{array}{c}
(\bar3,1)\otimes(3,1)\oplus(1,\bar3)\otimes(1,3)
\\ \hline (\bar3,1)\otimes(\bar6,1)\oplus(1,\bar3)\otimes(1,\bar6)
\\ \hline (6,1)\otimes(3,1)\oplus(1,6)\otimes(1,3)
\\ \hline (6,1)\otimes(\bar6,1)\oplus(1,6)\otimes(1,\bar6)
\end{array}$
\\ \hline
$q_L q_L \bar q_L \bar q_R + q_R q_R \bar q_R \bar q_L$
& $\begin{array}{c}
~~~[({\mathbf 3},\bar{\mathbf 3}) \oplus (\bar{\mathbf 3},{\mathbf 3})] \, , \, [(\bar{\mathbf 6},\bar{\mathbf 3}) \oplus (\bar{\mathbf 3},\bar{\mathbf 6})]~~~
\\ \hline [({\mathbf 3},\bar{\mathbf 3}) \oplus (\bar{\mathbf 3},{\mathbf 3})] \, , \, [({\mathbf {15}},\bar{\mathbf 3}) \oplus (\bar{\mathbf 3},{\mathbf {15}})]
\end{array}$
& $\begin{array}{c}
(\bar3,1)\otimes(\bar3,\bar3)\oplus(1,\bar3)\otimes(\bar3,\bar3)
\\ \hline (6,1)\otimes(\bar3,\bar3)\oplus(1,6)\otimes(\bar3,\bar3)
\end{array}$
\\ \hline
$q_L q_L \bar q_R \bar q_R + q_R q_R \bar q_L \bar q_L$
& $\begin{array}{c}
~~~~~~~~~~~~~[(\bar{\mathbf 3},{\mathbf 3}) \oplus ({\mathbf 3},\bar{\mathbf 3})]~~~~~~~~~~~~~
\\ \hline [(\bar{\mathbf 3},\bar{\mathbf 6}) \oplus (\bar{\mathbf 6},\bar{\mathbf 3})]
\\ \hline [({\mathbf 6},{\mathbf 3}) \oplus ({\mathbf 3},{\mathbf 6})]
\\ \hline [({\mathbf 6},\bar{\mathbf 6}) \oplus (\bar{\mathbf 6},{\mathbf 6})]
\end{array}$
& $\begin{array}{c}
(\bar3,1)\otimes(1,3)\oplus(1,\bar3)\otimes(3,1)
\\ \hline (\bar3,1)\otimes(1,\bar6)\oplus(1,\bar3)\otimes(\bar6,1)
\\ \hline (6,1)\otimes(1,3)\oplus(1,6)\otimes(3,1)
\\ \hline (6,1)\otimes(1,\bar6)\oplus(1,6)\otimes(\bar6,1)
\end{array}$
\\ \hline
$q_L q_R \bar q_L \bar q_L + q_R q_L \bar q_R \bar q_R$
& $\begin{array}{c}
~~~[(\bar{\mathbf 3},{\mathbf 3}) \oplus ({\mathbf 3},\bar{\mathbf 3})] \, , \, [({\mathbf 6},{\mathbf 3}) \oplus ({\mathbf 3},{\mathbf 6})]~~~
\\ \hline [(\bar{\mathbf 3},{\mathbf 3}) \oplus ({\mathbf 3},\bar{\mathbf 3})] \, , \, [(\overline{\mathbf {15}},{\mathbf 3}) \oplus ({\mathbf 3},\overline{\mathbf {15}})]
\end{array}$
& $\begin{array}{c}
(3,3)\otimes(3,1)\oplus(3,3)\otimes(1,3)
\\ \hline (3,3)\otimes(\bar6,1)\oplus(3,3)\otimes(1,\bar6)
\end{array}$
\\ \hline
$q_L q_R \bar q_L \bar q_R + q_R q_L \bar q_R \bar q_L$
& $\begin{array}{c}
~~~[({\mathbf 1},{\mathbf 1}) \oplus ({\mathbf 1},{\mathbf 1})] \, , \, [({\mathbf 8},{\mathbf 1}) \oplus ({\mathbf 1},{\mathbf 8})]~~~
\\ \hline [({\mathbf 1},{\mathbf 8}) \oplus ({\mathbf 8},{\mathbf 1})] \, , \, [({\mathbf 8},{\mathbf 8}) \oplus ({\mathbf 8},{\mathbf 8})]
\end{array}$
& $(3,3)\otimes(\bar3,\bar3)\oplus(3,3)\otimes(\bar3,\bar3)$
\\ \hline\hline
\end{tabular}
\label{tab:chiral}
\end{center}
\renewcommand{\arraystretch}{1}
\end{table}

Using the same definitions and notations shown in the previous section, we can explicitly write down these chiral representations. Their explicit forms are shown in several tables. The flavor singlet tetraquarks are shown in Tab~\ref{tab:singlet}, the flavor decuplet, anti-decuplet and $\mathbf{27}_f$ tetraquarks are shown in Tab~\ref{tab:decuplet}, and the flavor octet tetraquarks are shown in Tab~\ref{tab:octet}. In these tables there are many exotic chiral multiplets, such as the $[(\overline {\mathbf {15}}, {\mathbf 3})\oplus({\mathbf 3}, \overline {\mathbf {15}})]$ chiral multiplet. In this paper we shall not study these multiplets, but concentrate on the ``non-exotic'' $[(\bar \mathbf{3}, \mathbf{3})\oplus(\mathbf{3}, \bar \mathbf{3})]$ and $[({\mathbf 8},{\mathbf 1}) \oplus ({\mathbf 1},{\mathbf 8})]$ chiral multiplets as well as their mirror multiplets, which have the same chiral representations as the lowest level $\bar q q$ mesons.

\begin{table}[!hbt]
\renewcommand{\arraystretch}{1.5}
\begin{center}
\caption{The flavor singlet tetraquarks and their chiral structures.}
\begin{tabular}{c c c c}
\hline\hline
Tetraquarks & Total Expression & Chiral Rep. & Diquark Symmetry ($q q \otimes \bar q \bar q$)
\\ \hline \hline
$\mathcal{S}^{({\mathbf 1}, {\mathbf 1})}$ & $\epsilon^{ABE} \epsilon^{CDE} \big ( q_L^A q_L^B \bar q_L^C \bar q_L^D + q_R^A q_R^B \bar q_R^C \bar q_R^D \big )$
& $[({\mathbf 1}, {\mathbf 1}) + ({\mathbf 1}, {\mathbf 1})]$ & $(\bar3,1)\otimes(3,1)\oplus(1,\bar3)\otimes(1,3)$
\\ \hline
$\mathcal{S}^{({\mathbf 1}, {\mathbf 1})}$ & $\big (\delta^{AC} \delta^{BD} + \delta^{AD} \delta^{BC} \big) \big ( q_L^A q_L^B \bar q_L^C \bar q_L^D + q_R^A q_R^B \bar q_R^C \bar q_R^D \big )$
& $[({\mathbf 1}, {\mathbf 1}) + ({\mathbf 1}, {\mathbf 1})]$ & $(6,1)\otimes(\bar6,1)\oplus(1,6)\otimes(1,\bar6)$
\\ \hline
$\mathcal{S}^{({\mathbf 1}, {\mathbf 1})}$ & $\delta^{AC} \delta^{BD} \big ( q_L^A q_R^B \bar q_L^C \bar q_R^D + q_R^A q_L^B \bar q_R^C \bar q_L^D \big )$
& $[({\mathbf 1}, {\mathbf 1}) + ({\mathbf 1}, {\mathbf 1})]$ & $(3,3)\otimes(\bar3,\bar3)\oplus(3,3)\otimes(\bar3,\bar3)$
\\ \hline
$\mathcal{S}^{({\mathbf 8}, {\mathbf 8})}$ & $\lambda_N^{AC} \lambda_N^{BD} \big ( q_L^A q_R^B \bar q_L^C \bar q_R^D + q_R^A q_L^B \bar q_R^C \bar q_L^D \big )$
& $[({\mathbf 8}, {\mathbf 8}) + ({\mathbf 8}, {\mathbf 8})]$ & $(3,3)\otimes(\bar3,\bar3)\oplus(3,3)\otimes(\bar3,\bar3)$
\\ \hline
$\mathcal{S}^{({\mathbf 3}, \bar {\mathbf 3})}$ & $\epsilon^{ABE} \epsilon^{CDE} \big ( q_L^A q_L^B \bar q_L^C \bar q_R^D + q_R^A q_R^B \bar q_R^C \bar q_L^D \big )$
& $[({\mathbf 3}, \bar {\mathbf 3}) + (\bar {\mathbf 3}, {\mathbf 3})]$ & $(\bar3,1)\otimes(\bar3,\bar3)\oplus(1,\bar3)\otimes(\bar3,\bar3)$
\\ \hline
$\mathcal{S}^{({\mathbf 3}, \bar {\mathbf 3})}$ & $\big (\delta^{AC} \delta^{BD} + \delta^{BC} \delta^{AD} \big) \big ( q_L^A q_L^B \bar q_L^C \bar q_R^D + q_R^A q_R^B \bar q_R^C \bar q_L^D \big )$
& $[({\mathbf 3}, \bar {\mathbf 3}) + (\bar {\mathbf 3}, {\mathbf 3})]$ & $(6,1)\otimes(\bar3,\bar3)\oplus(1,6)\otimes(\bar3,\bar3)$
\\ \hline
$\mathcal{S}^{(\bar {\mathbf 3}, {\mathbf 3})}$ & $\epsilon^{ABE} \epsilon^{CDE} \big ( q_L^A q_L^B \bar q_R^C \bar q_R^D + q_R^A q_R^B \bar q_L^C \bar q_L^D \big )$
& $[(\bar {\mathbf 3}, {\mathbf 3}) + ({\mathbf 3}, \bar {\mathbf 3})]$ & $(\bar3,1)\otimes(1,3)\oplus(1,\bar3)\otimes(3,1)$
\\ \hline
$\mathcal{S}^{(\bar {\mathbf 3}, {\mathbf 3})}$ & $\epsilon^{ABE} \epsilon^{CDE} \big ( q_L^A q_R^B \bar q_L^C \bar q_L^D + q_R^A q_L^B \bar q_R^C \bar q_R^D \big )$
& $[(\bar {\mathbf 3}, {\mathbf 3}) + ({\mathbf 3}, \bar {\mathbf 3})]$ & $(3,3)\otimes(3,1)\oplus(3,3)\otimes(1,3)$
\\ \hline
$\mathcal{S}^{(\bar {\mathbf 3}, {\mathbf 3})}$ & $\big (\delta^{AC} \delta^{BD} + \delta^{AD} \delta^{BC} \big) \big ( q_L^A q_R^B \bar q_L^C \bar q_L^D + q_R^A q_L^B \bar q_R^C \bar q_R^D \big )$
& $[(\bar {\mathbf 3}, {\mathbf 3}) + ({\mathbf 3}, \bar {\mathbf 3})]$ & $(3,3)\otimes(\bar6,1)\oplus(3,3)\otimes(1,\bar6)$
\\ \hline
$\mathcal{S}^{({\mathbf 6}, \bar {\mathbf 6})}$ & $\big (\delta^{AC} \delta^{BD} + \delta^{AD} \delta^{BC} \big) \big ( q_L^A q_L^B \bar q_R^C \bar q_R^D + q_R^A q_R^B \bar q_L^C \bar q_L^D \big )$
& $[({\mathbf 6}, \bar {\mathbf 6}) + (\bar {\mathbf 6}, {\mathbf 6})]$ & $(6,1)\otimes(1,\bar6)\oplus(1,6)\otimes(\bar6,1)$
\\ \hline\hline
\end{tabular}
\label{tab:singlet}
\end{center}
\renewcommand{\arraystretch}{1}
\end{table}

\begin{table}[!hbt]
\renewcommand{\arraystretch}{1.5}
\begin{center}
\caption{The flavor decuplet, anti-decuplet and $\mathbf{27}_f$ tetraquarks and their chiral structures.}
\begin{tabular}{c c c c}
\hline\hline
Tetraquarks & Total Expression & Chiral Rep. & Diquark Symmetry ($q q \otimes \bar q \bar q$)
\\ \hline\hline
$\mathcal{D}^{({\mathbf {10}}, {\mathbf 1})}$ & $S_P^{ABE} \epsilon^{CDE} \big ( q_L^A q_L^B \bar q_L^C \bar q_L^D + q_R^A q_R^B \bar q_R^C \bar q_R^D \big )$
& $[({\mathbf {10}}, {\mathbf 1}) + ({\mathbf 1}, {\mathbf {10}})]$ & $(6,1)\otimes(3,1)\oplus(1,6)\otimes(1,3)$
\\ \hline
$\mathcal{D}^{({\mathbf {15}}, \bar {\mathbf 3})}$ & $S_P^{ABE} \epsilon^{CDE} \big ( q_L^A q_L^B \bar q_L^C \bar q_R^D + q_R^A q_R^B \bar q_R^C \bar q_L^D \big )$
& $[({\mathbf {15}}, \bar {\mathbf 3}) + (\bar {\mathbf 3}, {\mathbf {15}})]$ & $(6,1)\otimes(\bar3,\bar3)\oplus(1,6)\otimes(\bar3,\bar3)$
\\ \hline
$\mathcal{D}^{({\mathbf 6}, {\mathbf 3})}$ & $S_P^{ABE} \epsilon^{CDE} \big ( q_L^A q_L^B \bar q_R^C \bar q_R^D + q_R^A q_R^B \bar q_L^C \bar q_L^D \big )$
& $[({\mathbf 6}, {\mathbf 3}) + ({\mathbf 3}, {\mathbf 6})]$ & $(6,1)\otimes(1,3)\oplus(1,6)\otimes(3,1)$
\\ \hline
$\mathcal{D}^{({\mathbf 6}, {\mathbf 3})}$ & $S_P^{ABE} \epsilon^{CDE} \big ( q_L^A q_R^B \bar q_L^C \bar q_L^D + q_R^A q_L^B \bar q_R^C \bar q_R^D \big )$
& $[({\mathbf 6}, {\mathbf 3}) + ({\mathbf 3}, {\mathbf 6})]$ & $(3,3)\otimes(3,1)\oplus(3,3)\otimes(1,3)$
\\ \hline
$\mathcal{D}^{({\mathbf 8}, {\mathbf 8})}$ & $S_P^{ABE} \epsilon^{CDE} \big ( q_L^A q_R^B \bar q_L^C \bar q_R^D + q_R^A q_L^B \bar q_R^C \bar q_L^D \big )$
& $[({\mathbf 8}, {\mathbf 8}) + ({\mathbf 8}, {\mathbf 8})]$ & $(3,3)\otimes(\bar3,\bar3)\oplus(3,3)\otimes(\bar3,\bar3)$
\\ \hline\hline
$\overline\mathcal{D}^{(\overline {\mathbf {10}}, {\mathbf 1})}$ & $\epsilon^{ABE} S_P^{CDE} \big ( q_L^A q_L^B \bar q_L^C \bar q_L^D + q_R^A q_R^B \bar q_R^C \bar q_R^D \big )$
& $[(\overline {\mathbf {10}}, {\mathbf 1}) + ({\mathbf 1}, \overline {\mathbf {10}})]$ & $(\bar3,1)\otimes(\bar6,1)\oplus(1,\bar3)\otimes(1,\bar6)$
\\ \hline
$\overline\mathcal{D}^{(\bar{\mathbf 6}, \bar {\mathbf 3})}$ & $\epsilon^{ABE} S_P^{CDE} \big ( q_L^A q_L^B \bar q_L^C \bar q_R^D + q_R^A q_R^B \bar q_R^C \bar q_L^D \big )$
& $[(\bar{\mathbf 6}, \bar {\mathbf 3}) + (\bar {\mathbf 3}, \bar {\mathbf 6})]$ & $(\bar3,1)\otimes(\bar3,\bar3)\oplus(1,\bar3)\otimes(\bar3,\bar3)$
\\ \hline
$\overline\mathcal{D}^{(\bar{\mathbf 3}, \bar {\mathbf 6})}$ & $\epsilon^{ABE} S_P^{CDE} \big ( q_L^A q_L^B \bar q_R^C \bar q_R^D + q_R^A q_R^B \bar q_L^C \bar q_L^D \big )$
& $[(\bar{\mathbf 3}, \bar{\mathbf 6}) + (\bar{\mathbf 6}, \bar{\mathbf 3})]$ & $(\bar3,1)\otimes(1,\bar6)\oplus(1,\bar3)\otimes(\bar6,1)$
\\ \hline
$\overline\mathcal{D}^{(\overline {\mathbf {15}}, {\mathbf 3})}$ & $\epsilon^{ABE} S_P^{CDE} \big ( q_L^A q_R^B \bar q_L^C \bar q_L^D + q_R^A q_L^B \bar q_R^C \bar q_R^D \big )$
& $[(\overline {\mathbf {15}}, {\mathbf 3}) + ({\mathbf 3}, \overline {\mathbf {15}})]$ & $(3,3)\otimes(\bar6,1)\oplus(3,3)\otimes(1,\bar6)$
\\ \hline
$\overline\mathcal{D}^{({\mathbf 8}, {\mathbf 8})}$ & $\epsilon^{ABE} S_P^{CDE} \big ( q_L^A q_R^B \bar q_L^C \bar q_R^D + q_R^A q_L^B \bar q_R^C \bar q_L^D \big )$
& $[({\mathbf 8}, {\mathbf 8}) + ({\mathbf 8}, {\mathbf 8})]$ & $(3,3)\otimes(\bar3,\bar3)\oplus(3,3)\otimes(\bar3,\bar3)$
\\ \hline \hline
$\mathcal{T}^{({\mathbf {27}}, {\mathbf 1})}$ & $S_U^{ABCD} \big ( q_L^A q_L^B \bar q_L^C \bar q_L^D + q_R^A q_R^B \bar q_R^C \bar q_R^D \big )$
& $[({\mathbf {27}}, {\mathbf 1}) + ({\mathbf 1}, {\mathbf {27}})]$ & $(6,1)\otimes(\bar 6,1)\oplus(1,6)\otimes(1,\bar 6)$
\\ \hline
$\mathcal{T}^{({\mathbf {15}}, \bar {\mathbf 3})}$ & $S_U^{ABCD} \big ( q_L^A q_L^B \bar q_L^C \bar q_R^D + q_R^A q_R^B \bar q_R^C \bar q_L^D \big )$
& $[({\mathbf {15}}, \bar {\mathbf 3}) + (\bar {\mathbf 3}, {\mathbf {15}})]$ & $(6,1)\otimes(\bar3,\bar3) \oplus (1,6)\otimes(\bar3,\bar3)$
\\ \hline
$\mathcal{T}^{(\overline {\mathbf {15}}, {\mathbf 3})}$ & $S_U^{ABCD} \big ( q_L^A q_R^B \bar q_L^C \bar q_L^D + q_R^A q_L^B \bar q_R^C \bar q_R^D \big )$
& $[(\overline {\mathbf {15}}, {\mathbf 3}) + ({\mathbf 3}, \overline {\mathbf {15}})]$ & $(3,3)\otimes(\bar6,1) \oplus (3,3)\otimes(1,\bar6)$
\\ \hline
$\mathcal{T}^{({\mathbf 8}, {\mathbf 8})}$ & $S_U^{ABCD} \big ( q_L^A q_R^B \bar q_L^C \bar q_R^D + q_R^A q_L^B \bar q_R^C \bar q_L^D \big )$
& $[({\mathbf 8}, {\mathbf 8}) + ({\mathbf 8}, {\mathbf 8})]$ & $(3,3)\otimes(\bar3,\bar3) \oplus (3,3)\otimes(\bar3,\bar3)$
\\ \hline \hline
\end{tabular}
\label{tab:decuplet}
\end{center}
\renewcommand{\arraystretch}{1}
\end{table}

\begin{table}[!hbt]
\tiny
\begin{center}
\caption{The flavor octet tetraquarks and their chiral structures.}
\begin{tabular}{c c c c}
\hline\hline
Tetraquarks & Total Expression & Chiral Rep. & Diquark Symmetry ($q q \otimes \bar q \bar q$)
\\ \hline \hline
$\mathcal{O}^{({\mathbf 8}, {\mathbf 1})}$ & $ \big ( \delta^{AC} \lambda_N^{DB} - \delta^{BC} \lambda_N^{DA} - \delta^{AD} \lambda_N^{CB} + \delta^{BD} \lambda_N^{CA} \big ) \big ( q_L^A q_L^B \bar q_L^C \bar q_L^D + q_R^A q_R^B \bar q_R^C \bar q_R^D \big )$
& $[({\mathbf 8}, {\mathbf 1}) + ({\mathbf 1}, {\mathbf 8})]$ & $(\bar3,1)\otimes(3,1)\oplus(1,\bar3)\otimes(1,3)$
\\ \hline
$\mathcal{O}^{({\mathbf 8}, {\mathbf 1})}$ & $ \big ( \delta^{AC} \lambda_N^{DB} - \delta^{BC} \lambda_N^{DA} + \delta^{AD} \lambda_N^{CB} - \delta^{BD} \lambda_N^{CA} \big ) \big ( q_L^A q_L^B \bar q_L^C \bar q_L^D + q_R^A q_R^B \bar q_R^C \bar q_R^D \big )$
& $[({\mathbf 8}, {\mathbf 1}) + ({\mathbf 1}, {\mathbf 8})]$ & $(\bar3,1)\otimes(\bar6,1)\oplus(1,\bar3)\otimes(1,\bar6)$
\\ \hline
$\mathcal{O}^{({\mathbf 8}, {\mathbf 1})}$ & $ \big ( \delta^{AC} \lambda_N^{DB} + \delta^{BC} \lambda_N^{DA} - \delta^{AD} \lambda_N^{CB} - \delta^{BD} \lambda_N^{CA} \big ) \big ( q_L^A q_L^B \bar q_L^C \bar q_L^D + q_R^A q_R^B \bar q_R^C \bar q_R^D \big )$
& $[({\mathbf 8}, {\mathbf 1}) + ({\mathbf 1}, {\mathbf 8})]$ & $(6,1)\otimes(3,1)\oplus(1,6)\otimes(1,3)$
\\ \hline
$\mathcal{O}^{({\mathbf 8}, {\mathbf 1})}$ & $ \big ( \delta^{AC} \lambda_N^{DB} + \delta^{BC} \lambda_N^{DA} + \delta^{AD} \lambda_N^{CB} + \delta^{BD} \lambda_N^{CA} \big ) \big ( q_L^A q_L^B \bar q_L^C \bar q_L^D + q_R^A q_R^B \bar q_R^C \bar q_R^D \big )$
& $[({\mathbf 8}, {\mathbf 1}) + ({\mathbf 1}, {\mathbf 8})]$ & $(6,1)\otimes(\bar6,1)\oplus(1,6)\otimes(1,\bar6)$
\\ \hline
$\mathcal{O}^{({\mathbf 3}, \bar {\mathbf 3})}$ & $ \big ( \delta^{AC} \lambda_N^{DB} - \delta^{BC} \lambda_N^{DA} \big ) \big ( q_L^A q_L^B \bar q_L^C \bar q_R^D + q_R^A q_R^B \bar q_R^C \bar q_L^D \big )$
& $[({\mathbf 3}, \bar {\mathbf 3}) + (\bar {\mathbf 3}, {\mathbf 3})]$ & $(\bar3,1)\otimes(\bar3,\bar3)\oplus(1,\bar3)\otimes(\bar3,\bar3)$
\\ \hline
$\mathcal{O}^{(\bar {\mathbf 6}, \bar {\mathbf 3})}$ & $ \big ( \delta^{AC} \lambda_N^{DB} - \delta^{BC} \lambda_N^{DA} - 2 \delta^{AD} \lambda_N^{CB} + 2 \delta^{BD} \lambda_N^{CA} \big ) \big ( q_L^A q_L^B \bar q_L^C \bar q_R^D + q_R^A q_R^B \bar q_R^C \bar q_L^D \big )$
& $[(\bar {\mathbf 6}, \bar {\mathbf 3}) + (\bar {\mathbf 3}, \bar {\mathbf 6})]$ & $(\bar3,1)\otimes(\bar3,\bar3)\oplus(1,\bar3)\otimes(\bar3,\bar3)$
\\ \hline
$\mathcal{O}^{({\mathbf 3}, \bar {\mathbf 3})}$ & $ \big ( \delta^{AC} \lambda_N^{DB} + \delta^{BC} \lambda_N^{DA} \big ) \big ( q_L^A q_L^B \bar q_L^C \bar q_R^D + q_R^A q_R^B \bar q_R^C \bar q_L^D \big )$
& $[({\mathbf 3}, \bar {\mathbf 3}) + (\bar {\mathbf 3}, {\mathbf 3})]$ & $(6,1)\otimes(\bar3,\bar3)\oplus(1,6)\otimes(\bar3,\bar3)$
\\ \hline
$\mathcal{O}^{({\mathbf {15}}, \bar {\mathbf 3})}$ & $ \big ( \delta^{AD} \lambda_N^{CB} + \delta^{BD} \lambda_N^{CA} \big ) \big ( q_L^A q_L^B \bar q_L^C \bar q_R^D + q_R^A q_R^B \bar q_R^C \bar q_L^D \big )$
& $[({\mathbf {15}}, \bar {\mathbf 3}) + (\bar {\mathbf 3}, {\mathbf {15}})]$ & $(6,1)\otimes(\bar3,\bar3)\oplus(1,6)\otimes(\bar3,\bar3)$
\\ \hline
$\mathcal{O}^{(\bar {\mathbf 3}, {\mathbf 3})}$ & $ \big ( \delta^{AC} \lambda_N^{DB} - \delta^{BC} \lambda_N^{DA} - \delta^{AD} \lambda_N^{CB} + \delta^{BD} \lambda_N^{CA} \big ) \big ( q_L^A q_L^B \bar q_R^C \bar q_R^D + q_R^A q_R^B \bar q_L^C \bar q_L^D \big )$
& $[(\bar {\mathbf 3}, {\mathbf 3}) + ({\mathbf 3}, \bar {\mathbf 3})]$ & $(\bar3,1)\otimes(1,3)\oplus(1,\bar3)\otimes(3,1)$
\\ \hline
$\mathcal{O}^{(\bar {\mathbf 3}, \bar {\mathbf 6})}$ & $ \big ( \delta^{AC} \lambda_N^{DB} - \delta^{BC} \lambda_N^{DA} + \delta^{AD} \lambda_N^{CB} - \delta^{BD} \lambda_N^{CA} \big ) \big ( q_L^A q_L^B \bar q_R^C \bar q_R^D + q_R^A q_R^B \bar q_L^C \bar q_L^D \big )$
& $[(\bar {\mathbf 3}, \bar {\mathbf 6}) + (\bar {\mathbf 6}, \bar {\mathbf 3})]$ & $(\bar3,1)\otimes(1,\bar6)\oplus(1,\bar3)\otimes(\bar6,1)$
\\ \hline
$\mathcal{O}^{({\mathbf 6}, {\mathbf 3})}$ & $ \big ( \delta^{AC} \lambda_N^{DB} + \delta^{BC} \lambda_N^{DA} - \delta^{AD} \lambda_N^{CB} - \delta^{BD} \lambda_N^{CA} \big ) \big ( q_L^A q_L^B \bar q_R^C \bar q_R^D + q_R^A q_R^B \bar q_L^C \bar q_L^D \big )$
& $[({\mathbf 6}, {\mathbf 3}) + ({\mathbf 3}, {\mathbf 6})]$ & $(6,1)\otimes(1,3)\oplus(1,6)\otimes(3,1)$
\\ \hline
$\mathcal{O}^{({\mathbf 6}, \bar {\mathbf 6})}$ & $ \big ( \delta^{AC} \lambda_N^{DB} + \delta^{BC} \lambda_N^{DA} + \delta^{AD} \lambda_N^{CB} + \delta^{BD} \lambda_N^{CA} \big ) \big ( q_L^A q_L^B \bar q_R^C \bar q_R^D + q_R^A q_R^B \bar q_L^C \bar q_L^D \big )$
& $[({\mathbf 6}, \bar {\mathbf 6}) + (\bar {\mathbf 6}, {\mathbf 6})]$ & $(6,1)\otimes(1,\bar6)\oplus(1,6)\otimes(\bar 6,1)$
\\ \hline
$\mathcal{O}^{(\bar {\mathbf 3}, {\mathbf 3})}$ & $ \big ( \delta^{AC} \lambda_N^{DB} - \delta^{AD} \lambda_N^{CB} \big ) \big ( q_L^A q_R^B \bar q_L^C \bar q_L^D + q_R^A q_L^B \bar q_R^C \bar q_R^D \big )$
& $[(\bar {\mathbf 3}, {\mathbf 3}) + ({\mathbf 3}, \bar {\mathbf 3})]$ & $(3,3)\otimes(3,1)\oplus(3,3)\otimes(1,3)$
\\ \hline
$\mathcal{O}^{({\mathbf 6}, {\mathbf 3})}$ & $ \big ( \delta^{AC} \lambda_N^{DB} - \delta^{AD} \lambda_N^{CB} - 2 \delta^{BC} \lambda_N^{DA} + 2 \delta^{BD} \lambda_N^{CA} \big ) \big ( q_L^A q_R^B \bar q_L^C \bar q_L^D + q_R^A q_L^B \bar q_R^C \bar q_R^D \big )$
& $[({\mathbf 6}, {\mathbf 3}) + ({\mathbf 3}, {\mathbf 6})]$ & $(3,3)\otimes(3,1)\oplus(3,3)\otimes(1,3)$
\\ \hline
$\mathcal{O}^{(\bar {\mathbf 3}, {\mathbf 3})}$ & $ \big ( \delta^{AC} \lambda_N^{DB} + \delta^{AD} \lambda_N^{CB} \big ) \big ( q_L^A q_R^B \bar q_L^C \bar q_L^D + q_R^A q_L^B \bar q_R^C \bar q_R^D \big )$
& $[(\bar {\mathbf 3}, {\mathbf 3}) + ({\mathbf 3}, \bar {\mathbf 3})]$ & $(3,3)\otimes(\bar6,1)\oplus(3,3)\otimes(1,\bar6)$
\\ \hline
$\mathcal{O}^{(\overline {\mathbf {15}}, {\mathbf 3})}$ & $ \big ( \delta^{BC} \lambda_N^{DA} + \delta^{BD} \lambda_N^{CA} \big ) \big ( q_L^A q_R^B \bar q_L^C \bar q_L^D + q_R^A q_L^B \bar q_R^C \bar q_R^D \big )$
& $[(\overline {\mathbf {15}}, {\mathbf 3}) + ({\mathbf 3}, \overline {\mathbf {15}})]$ & $(3,3)\otimes(\bar6,1)\oplus(3,3)\otimes(1,\bar6)$
\\ \hline
$\mathcal{O}^{({\mathbf 8}, {\mathbf 1})}$ & $ \delta^{BD} \lambda_N^{CA} \big ( q_L^A q_R^B \bar q_L^C \bar q_R^D + q_R^A q_L^B \bar q_R^C \bar q_L^D \big )$
& $[({\mathbf 8}, {\mathbf 1}) + ({\mathbf 1}, {\mathbf 8})]$ & $(3,3)\otimes(\bar3,\bar3)\oplus(3,3)\otimes(\bar3,\bar3)$
\\ \hline
$\mathcal{O}^{({\mathbf 1}, {\mathbf 8})}$ & $ \delta^{AC} \lambda_N^{DB} \big ( q_L^A q_R^B \bar q_L^C \bar q_R^D + q_R^A q_L^B \bar q_R^C \bar q_L^D \big )$
& $[({\mathbf 1}, {\mathbf 8}) + ({\mathbf 8}, {\mathbf 1})]$ & $(3,3)\otimes(\bar3,\bar3)\oplus(3,3)\otimes(\bar3,\bar3)$
\\ \hline
$\mathcal{O}^{({\mathbf 8}, {\mathbf 8})}$ & $ \delta^{AD} \lambda_N^{CB} \big ( q_L^A q_R^B \bar q_L^C \bar q_R^D + q_R^A q_L^B \bar q_R^C \bar q_L^D \big )$
& $[({\mathbf 8}, {\mathbf 8}) + ({\mathbf 8}, {\mathbf 8})]$ & $(3,3)\otimes(\bar3,\bar3)\oplus(3,3)\otimes(\bar3,\bar3)$
\\ \hline
$\mathcal{O}^{({\mathbf 8}, {\mathbf 8})}$ & $ \delta^{BC} \lambda_N^{DA} \big ( q_L^A q_R^B \bar q_L^C \bar q_R^D + q_R^A q_L^B \bar q_R^C \bar q_L^D \big )$
& $[({\mathbf 8}, {\mathbf 8}) + ({\mathbf 8}, {\mathbf 8})]$ & $(3,3)\otimes(\bar3,\bar3)\oplus(3,3)\otimes(\bar3,\bar3)$
\\ \hline\hline
\end{tabular}
\label{tab:octet}
\end{center}
\end{table}

From Tabs~\ref{tab:singlet} and \ref{tab:octet} we find that there are altogether two $[(\mathbf{3}, \bar \mathbf{3})\oplus(\bar \mathbf{3}, \mathbf{3})]$ chiral multiplets $\mathcal{T}^{(\mathbf{3}, \bar\mathbf{3})}_{1,2}$ and three $[(\bar \mathbf{3}, \mathbf{3})\oplus(\mathbf{3}, \bar \mathbf{3})]$ multiplets $\mathcal{T}^{(\bar \mathbf{3},\mathbf{3})}_{3,4,5}$. Each contains one singlet and one octet tetraquarks:
\begin{eqnarray}
\mathcal{T}^{(\mathbf{3}, \bar\mathbf{3})/(\bar\mathbf{3},\mathbf{3})}_i = (\mathcal{S}_i \, , \, \mathcal{O}_i)^T \, .
\end{eqnarray}
The singlet and octet tetraquarks contain both the left-handed and right-handed parts which can be written in the following way in order to specify representations of the chiral group $SU(3)_L \otimes SU(3)_R$ for simplicity, i.e., $\mathcal{T} = \mathcal{T}_L + \mathcal{T}_R \rightarrow \left ( \begin{array}{c} \mathcal{T}_L \\ \mathcal{T}_R \end{array} \right )$, similarly to $q = q_L + q_R \rightarrow \left ( \begin{array}{c} q_L \\ q_R \end{array} \right )$~\cite{Jido:2001nt}:
\begin{eqnarray}
\mathcal{S}_1 &=& \epsilon^{ABE} \epsilon^{CDE} \big [ q_L^A q_L^B \bar q_L^C \bar q_R^D + q_R^A q_R^B \bar q_R^C \bar q_L^D \big ]
\rightarrow \epsilon^{ABE} \epsilon^{CDE}
\left ( \begin{array}{c}
q_L^A q_L^B \bar q_L^C \bar q_R^D
\\ q_R^A q_R^B \bar q_R^C \bar q_L^D
\end{array} \right ) \, ,
\\ \mathcal{O}_1 &=& ( \delta^{AC} \lambda_N^{DB} - \delta^{BC} \lambda_N^{DA} )
\left ( \begin{array}{c}
q_L^A q_L^B \bar q_L^C \bar q_R^D
\\ q_R^A q_R^B \bar q_R^C \bar q_L^D
\end{array} \right ) \, ,
\\ \mathcal{S}_2 &=& ( \delta^{AC} \delta^{BD} + \delta^{BC} \delta^{AD} )
\left ( \begin{array}{c}
q_L^A q_L^B \bar q_L^C \bar q_R^D
\\ q_R^A q_R^B \bar q_R^C \bar q_L^D
\end{array} \right ) \, ,
\\ \mathcal{O}_2 &=& ( \delta^{AC} \lambda_N^{DB} + \delta^{BC} \lambda_N^{DA} )
\left ( \begin{array}{c}
q_L^A q_L^B \bar q_L^C \bar q_R^D
\\ q_R^A q_R^B \bar q_R^C \bar q_L^D
\end{array} \right ) \, ,
\\ \mathcal{S}_3 &=& \epsilon^{ABE} \epsilon^{CDE}
\left ( \begin{array}{c}
q_L^A q_L^B \bar q_R^C \bar q_R^D
\\ q_R^A q_R^B \bar q_L^C \bar q_L^D
\end{array} \right ) \, ,
\\ \mathcal{O}_3 &=& ( \delta^{AC} \lambda_N^{DB} - \delta^{BC} \lambda_N^{DA} - \delta^{AD} \lambda_N^{CB} + \delta^{BD} \lambda_N^{CA} )
\left ( \begin{array}{c}
q_L^A q_L^B \bar q_R^C \bar q_R^D
\\ q_R^A q_R^B \bar q_L^C \bar q_L^D
\end{array} \right ) \, ,
\\ \mathcal{S}_4 &=& \epsilon^{ABE} \epsilon^{CDE}
\left ( \begin{array}{c}
q_L^A q_R^B \bar q_L^C \bar q_L^D
\\ q_R^A q_L^B \bar q_R^C \bar q_R^D
\end{array} \right ) \, ,
\\ \mathcal{O}_4 &=& ( \delta^{AC} \lambda_N^{DB} - \delta^{AD} \lambda_N^{CB} )
\left ( \begin{array}{c}
q_L^A q_R^B \bar q_L^C \bar q_L^D
\\ q_R^A q_L^B \bar q_R^C \bar q_R^D
\end{array} \right ) \, ,
\\ \mathcal{S}_5 &=& ( \delta^{AC} \delta^{BD} + \delta^{AD} \delta^{BC} )
\left ( \begin{array}{c}
q_L^A q_R^B \bar q_L^C \bar q_L^D
\\ q_R^A q_L^B \bar q_R^C \bar q_R^D
\end{array} \right ) \, ,
\\ \mathcal{O}_5 &=& ( \delta^{AC} \lambda_N^{DB} + \delta^{AD} \lambda_N^{CB} )
\left ( \begin{array}{c}
q_L^A q_R^B \bar q_L^C \bar q_R^D
\\ q_R^A q_L^B \bar q_R^C \bar q_L^D
\end{array} \right ) \, .
\end{eqnarray}
Again we find that the ``left'' and ``right'' can be interchanged here.

The chiral multiplet $\mathcal{T}^{(\mathbf{3}, \bar\mathbf{3})/(\bar\mathbf{3},\mathbf{3})}_{1,2,4,5}$ all contain one pair of quark and antiquark which has the same chirality and can be combined to be a chiral singlet. This is just like adding one chiral singlet quark-antiquark pair ($\bar q^A_L q^A_L + \bar q^A_R q^A_R$) to the lowest level $\bar q q$ chiral multiplet belonging to the representations $[(\mathbf{3}, \bar \mathbf{3})\oplus(\bar \mathbf{3}, \mathbf{3})]$ and $[(\bar \mathbf{3}, \mathbf{3})\oplus(\mathbf{3}, \bar \mathbf{3})]$. In contrast, all the quark-antiquark pairs contained in $\mathcal{T}^{(\bar\mathbf{3},\mathbf{3})}_{3}$ have the opposite chirality, and so can not be chiral singlet. This makes $\mathcal{T}^{(\bar\mathbf{3},\mathbf{3})}_{3}$ much different from others.

We do the same procedures and find there are altogether five $[({\mathbf 8},{\mathbf 1}) \oplus ({\mathbf 1},{\mathbf 8})]$ chiral multiplets $\mathcal{T}^{(\mathbf{8}, \mathbf{1})}_{1\cdots5}$ and one $[({\mathbf 1},{\mathbf 8}) \oplus ({\mathbf 8},{\mathbf 1})]$ multiplet $\mathcal{T}^{(\mathbf{1}, \mathbf{8})}_{6}$:
\begin{eqnarray}
\mathcal{T}^{(\mathbf{8}, \mathbf{1})}_{1,N} &=& ( \delta^{AC} \lambda_N^{DB} - \delta^{BC} \lambda_N^{DA} - \delta^{AD} \lambda_N^{CB} + \delta^{BD} \lambda_N^{CA} )
\left ( \begin{array}{c}
q_L^A q_L^B \bar q_L^C \bar q_L^D
\\ q_R^A q_R^B \bar q_R^C \bar q_R^D
\end{array} \right ) \, ,
\\ \mathcal{T}^{(\mathbf{8}, \mathbf{1})}_{2,N} &=& ( \delta^{AC} \lambda_N^{DB} - \delta^{BC} \lambda_N^{DA} + \delta^{AD} \lambda_N^{CB} - \delta^{BD} \lambda_N^{CA} )
\left ( \begin{array}{c}
q_L^A q_L^B \bar q_L^C \bar q_L^D
\\ q_R^A q_R^B \bar q_R^C \bar q_R^D
\end{array} \right ) \, ,
\\ \mathcal{T}^{(\mathbf{8}, \mathbf{1})}_{3,N} &=& ( \delta^{AC} \lambda_N^{DB} + \delta^{BC} \lambda_N^{DA} - \delta^{AD} \lambda_N^{CB} - \delta^{BD} \lambda_N^{CA} )
\left ( \begin{array}{c}
q_L^A q_L^B \bar q_L^C \bar q_L^D
\\ q_R^A q_R^B \bar q_R^C \bar q_R^D
\end{array} \right ) \, ,
\\ \mathcal{T}^{(\mathbf{8}, \mathbf{1})}_{4,N} &=& ( \delta^{AC} \lambda_N^{DB} + \delta^{BC} \lambda_N^{DA} + \delta^{AD} \lambda_N^{CB} + \delta^{BD} \lambda_N^{CA} )
\left ( \begin{array}{c}
q_L^A q_L^B \bar q_L^C \bar q_L^D
\\ q_R^A q_R^B \bar q_R^C \bar q_R^D
\end{array} \right ) \, ,
\\ \mathcal{T}^{(\mathbf{8}, \mathbf{1})}_{5,N} &=& \delta^{BD} \lambda_N^{CA}
\left ( \begin{array}{c}
q_L^A q_R^B \bar q_L^C \bar q_R^D
\\ q_R^A q_L^B \bar q_R^C \bar q_L^D
\end{array} \right ) \, ,
\\ \mathcal{T}^{(\mathbf{1}, \mathbf{8})}_{6,N} &=& \delta^{AC} \lambda_N^{DB}
\left ( \begin{array}{c}
q_L^A q_R^B \bar q_L^C \bar q_R^D
\\ q_R^A q_L^B \bar q_R^C \bar q_L^D
\end{array} \right ) \, .
\end{eqnarray}
All these six multiplets contain one pair of quark and antiquark which has the same chirality and can be combined to be a chiral singlet.

\section{Chiral Transformation}
\label{sec:transformation}

\subsection{Chiral Transformation of Mesons}

Under the $U(1)_V$, $U(1)_A$, $SU(3)_V$ and $SU(3)_A$ chiral transformations, the quark, $q=\left(\begin{array}{c}q_L \\ q_R\end{array}\right)$, transforms as
\begin{eqnarray}
\nonumber
\bf{U(1)_{V}} &:& q \to \exp(i a^0) q  = q + \delta q \, ,
\\
\bf{SU(3)_V} &:& q \to \exp (i \vec \lambda \cdot \vec a ){q} = q + \delta^{\vec{a}} q \, ,
\\ \nonumber
\bf{U(1)_{A}} &:& q \to \exp(i \gamma_5 b^0) q = q + \delta_5 q \, ,
\\ \nonumber
\bf{SU(3)_A} &:& q \to \exp (i \gamma_{5} \vec \lambda \cdot \vec b){q} = q + \delta_5^{\vec{b}} q \, ,
\end{eqnarray}
where $\vec \lambda$ are the eight Gell-Mann matrices; $a^0$ is an infinitesimal parameter for the $U(1)_V$ transformation, $\vec{a}$ the octet of $SU(3)_V$ group parameters, $b^0$ an infinitesimal parameter for the $U(1)_A$ transformation, and $\vec{b}$ the octet of the chiral transformations; the $\gamma_5$ matrix is diagonal in the chiral (Weyl) representations~\cite{Jido:2001nt}:
\begin{eqnarray}
\gamma_5 = \left ( \begin{array}{cc}
-1 & 0
\\ 0 & 1
\end{array} \right ) \, .
\end{eqnarray}
In this subsection we study the chiral transformation properties of quark-antiquark mesons. The chiral (flavor) structure of quark-antiquark mesons can be
\begin{eqnarray}
\mathcal{M}^{(\mathbf{1},\mathbf{1})}_1 &=& q^A_L \bar q^A_L + q^A_R \bar q^A_R \rightarrow \delta^{AB} \left ( \begin{array}{c} q^A_L \bar q^B_L \\ q^A_R \bar q^B_R \end{array} \right ) \, ,
\\ \mathcal{M}^{(\mathbf{8},\mathbf{1})}_{2,N} &=& \lambda_N^{BA} \left ( \begin{array}{c} q^A_L \bar q^B_L \\ q^A_R \bar q^B_R \end{array} \right ) \, ,
\\ \mathcal{M}^{(\mathbf{3},\mathbf{\bar3})}_3 &=& (\mathcal{S} \, , \, \mathcal{O})^T \, ,
\end{eqnarray}
where
\begin{eqnarray}
\mathcal{S} &=& \delta^{AB} \left ( \begin{array}{c} q^A_L \bar q^B_R \\ q^A_R \bar q^B_L \end{array} \right ) \, ,
\\ \mathcal{O}_N &=& \lambda_N^{BA} \left ( \begin{array}{c} q^A_L \bar q^B_R \\ q^A_R \bar q^B_L \end{array} \right ) \, .
\end{eqnarray}
The mesons $\mathcal{M}^{(\mathbf{1},\mathbf{1})}_1$, $\mathcal{M}^{(\mathbf{8},\mathbf{1})}_{2,N}$ and $\mathcal{M}^{(\mathbf{3},\mathbf{\bar3})}_3$ belong to the chiral representations $[(\mathbf{1},\mathbf{1})\oplus(\mathbf{1},\mathbf{1})]$, $[(\mathbf{8},\mathbf{1})\oplus(\mathbf{1},\mathbf{8})]$ and $[(\mathbf{3},\mathbf{\bar3})\oplus(\mathbf{\bar3},\mathbf{3})]$, respectively.

The $U(1)_V$ chiral transformation is trivial which counts the quark number $\delta \mathcal{M}_i = 0$. The other chiral transformation equations are:
\begin{eqnarray}
&& \left\{ \begin{array}{l}
\delta_5 \mathcal{M}^{(\mathbf{1},\mathbf{1})}_1 = 0 \, ,
\\ \delta^{\vec a} \mathcal{M}^{(\mathbf{1},\mathbf{1})}_1 = 0 \, ,
\\ \delta^{\vec b}_5 \mathcal{M}^{(\mathbf{1},\mathbf{1})}_1 = 0 \, .
\end{array} \right .
\label{eq:meson11}
\\ && \left\{ \begin{array}{l}
\delta_5 \mathcal{M}^{(\mathbf{8},\mathbf{1})}_{2,M} = 0 \, ,
\\ \delta^{\vec a} \mathcal{M}^{(\mathbf{8},\mathbf{1})}_{2,M} = 2 a^N f_{NMO} \mathcal{M}^{(\mathbf{8},\mathbf{1})}_{2,O} \, ,
\\ \delta^{\vec b}_5 \mathcal{M}^{(\mathbf{8},\mathbf{1})}_{2,M} = 2 i b^N f_{NMO} \gamma_5 \mathcal{M}^{(\mathbf{8},\mathbf{1})}_{2,O} \, .
\end{array} \right .
\label{eq:meson81}
\\ && \left\{ \begin{array}{l}
\delta_5 \mathcal{S} = 2 i b^0 \gamma_5 \mathcal{S} \, ,
\\ \delta^{\vec a} \mathcal{S} = 0 \, ,
\\ \delta^{\vec b}_5 \mathcal{S} = 2 i \gamma_5 b^N \mathcal{O}_N \, .
\end{array} \right .
\label{eq:meson331}
\\ && \left\{ \begin{array}{l}
\delta_5 \mathcal{O}_M = 2 i b^0 \gamma_5 \mathcal{O}_M \, ,
\\ \delta^{\vec a} \mathcal{O}_M = 2 a^N f_{NMO} \mathcal{O}_O \, ,
\\ \delta^{\vec b}_5 \mathcal{O}_M = 2 i b^N d_{NMO} \gamma_5 \mathcal{O}_O + {4\over3} b^M \gamma_5 \mathcal{S} \, .
\end{array} \right .
\label{eq:meson338}
\end{eqnarray}
We note that these chiral transformation equations are different from those of the relevant meson currents (fields), but their coefficients are similar. For example, the singlet meson field $\eta \equiv \bar q_A^a \gamma_5 q_A^a$, belonging to the chiral representation $[(\mathbf{3},\mathbf{\bar3})\oplus(\mathbf{\bar3},\mathbf{3})]$, transforms as
\begin{eqnarray}
\left\{ \begin{array}{l}
\delta_5 \eta = 2 i b^0 \bar q_A^a q_A^a \, ,
\\ \delta^{\vec a} \eta = 0 \, ,
\\ \delta^{\vec b}_5 \eta = 2 i b^N \bar q_A^a \lambda_N^{AB} q_B^a \, ,
\end{array} \right .
\end{eqnarray}
where the coefficients $2 i b^0$, $0$ and $2 i b^N$ are similar to those of Eqs.~(\ref{eq:meson331}).

\subsection{Chiral Transformation of Tetraquarks}

In this subsection we study the chiral transformation properties of tetraquarks belonging to the ``non-exotic'' $[(\bar \mathbf{3}, \mathbf{3})\oplus(\mathbf{3}, \bar \mathbf{3})]$ and $[({\mathbf 8},{\mathbf 1}) \oplus ({\mathbf 1},{\mathbf 8})]$ chiral representations. The calculations are straightforward and so we only show the final results here. The $U(1)_V$ chiral transformation is again trivial which counts the quark number $\delta \mathcal{T}_i = 0$. The chiral transformations of $\mathcal{T}^{(\mathbf{3},\bar\mathbf{3})}_{1,2}$ are the same, while $\mathcal{T}^{(\bar\mathbf{3},\mathbf{3})}_{4,5}$ transform like their mirror fields:
\begin{eqnarray}
&& \left\{ \begin{array}{l}
\delta_5 \mathcal{S}_1 = 2 i b^0 \gamma_5 \mathcal{S}_1 \, ,
\\ \delta^{\vec a} \mathcal{S}_1 = 0 \, ,
\\ \delta^{\vec b}_5 \mathcal{S}_1 = 2 i \gamma_5 b^N \mathcal{O}_1^N \, .
\end{array} \right .
\label{eq:tetraquark331}
\\ && \left\{ \begin{array}{l}
\delta_5 \mathcal{O}_1^M = 2 i b^0 \gamma_5 \mathcal{O}_1^M \, ,
\\ \delta^{\vec a} \mathcal{O}_1^M = 2 a^N f_{NMO} \mathcal{O}_1^O \, ,
\\ \delta^{\vec b}_5 \mathcal{O}_1^M = 2 i b^N d_{NMO} \gamma_5 \mathcal{O}_1^O + {4\over3} b^M \gamma_5 \mathcal{S}_1 \, .
\end{array} \right .
\\ && \left\{ \begin{array}{l}
\delta_5 \mathcal{S}_2 = 2 i b^0 \gamma_5 \mathcal{S}_2 \, ,
\\ \delta^{\vec a} \mathcal{S}_2 = 0 \, ,
\\ \delta^{\vec b}_5 \mathcal{S}_2 = 2 i \gamma_5 b^N \mathcal{O}_2^N \, .
\end{array} \right .
\\ && \left\{ \begin{array}{l}
\delta_5 \mathcal{O}_2^M = 2 i b^0 \gamma_5 \mathcal{O}_2^M \, ,
\\ \delta^{\vec a} \mathcal{O}_2^M = 2 a^N f_{NMO} \mathcal{O}_2^O \, ,
\\ \delta^{\vec b}_5 \mathcal{O}_2^M = 2 i b^N d_{NMO} \gamma_5 \mathcal{O}_2^O + {4\over3} b^M \gamma_5 \mathcal{S}_2 \, .
\end{array} \right .
\\ && \left\{ \begin{array}{l}
\delta_5 \mathcal{S}_4 = -2 i b^0 \gamma_5 \mathcal{S}_4 \, ,
\\ \delta^{\vec a} \mathcal{S}_4 = 0 \, ,
\\ \delta^{\vec b}_5 \mathcal{S}_4 = - 2 i \gamma_5 b^N \mathcal{O}_4^N \, .
\end{array} \right .
\\ && \left\{ \begin{array}{l}
\delta_5 \mathcal{O}_4^M = - 2 i b^0 \gamma_5 \mathcal{O}_4^M \, ,
\\ \delta^{\vec a} \mathcal{O}_4^M = 2 a^N f_{NMO} \mathcal{O}_4^O \, ,
\\ \delta^{\vec b}_5 \mathcal{O}_4^M = - 2 i b^N d_{NMO} \gamma_5 \mathcal{O}_4^O - {4\over3} b^M \gamma_5 \mathcal{S}_4 \, .
\end{array} \right .
\\ && \left\{ \begin{array}{l}
\delta_5 \mathcal{S}_5 = -2 i b^0 \gamma_5 \mathcal{S}_5 \, ,
\\ \delta^{\vec a} \mathcal{S}_5 = 0 \, ,
\\ \delta^{\vec b}_5 \mathcal{S}_5 = - 2 i \gamma_5 b^N \mathcal{O}_5^N \, .
\end{array} \right .
\\ && \left\{ \begin{array}{l}
\delta_5 \mathcal{O}_5^M = - 2 i b^0 \gamma_5 \mathcal{O}_5^M \, ,
\\ \delta^{\vec a} \mathcal{O}_5^M = 2 a^N f_{NMO} \mathcal{O}_5^O \, ,
\\ \delta^{\vec b}_5 \mathcal{O}_5^M = - 2 i b^N d_{NMO} \gamma_5 \mathcal{O}_5^O - {4\over3} b^M \gamma_5 \mathcal{S}_5 \, .
\end{array} \right .
\end{eqnarray}
These chiral transformation equations are the same as those of $(\sigma, \pi)$ belonging to the $[(\bar \mathbf{3}, \mathbf{3})\oplus(\mathbf{3}, \bar \mathbf{3})]$ chiral representation, or its mirror representation (Eqs.~(\ref{eq:meson331}) and (\ref{eq:meson338})). While the chiral transformation equations of $\mathcal{T}^{(\bar\mathbf{3},\mathbf{3})}_{3}$ are different
\begin{eqnarray}
&& \left\{ \begin{array}{l}
\delta_5 \mathcal{S}_3 = 4 i b^0 \gamma_5 \mathcal{S}_3 \, ,
\\ \delta^{\vec a} \mathcal{S}_3 = 0 \, ,
\\ \delta^{\vec b}_5 \mathcal{S}_3 = 2 i \gamma_5 b^N \mathcal{O}_3^N \, .
\end{array} \right .
\label{eq:tetraquark333}
\\ &&
\left\{ \begin{array}{l}
\delta_5 \mathcal{O}_3^M = 4 i b^0 \gamma_5 \mathcal{O}_3^M \, ,
\\ \delta^{\vec a} \mathcal{O}_3^M = 2 a^N f_{NMO} \mathcal{O}_3^O \, ,
\\ \delta^{\vec b}_5 \mathcal{O}_3^M = - 2 i b^N d_{NMO} \gamma_5 \mathcal{O}_3^O + {4\over3} b^M \gamma_5 \mathcal{S}_3 \, .
\end{array} \right .
\label{eq:tetraquark338}
\end{eqnarray}
We note that this is the only case which has a non-exotic chiral representation but transforms differently from others under chiral transformations. To obtain these equations we need to use the following relation:
\begin{eqnarray}
&& \lambda_N^{AC} \lambda_M^{BD}
\\ \nonumber &=& -{1\over12} \delta_{NM} \delta^{AC} \delta^{BD} + {1\over4} \delta_{NM} \delta^{AD} \delta^{BC}
\\ \nonumber && - {2\over5} d_{NMO} \delta^{AC} \lambda_O^{BD} + ({3\over5} d_{NMO} - {i\over3} f_{NMO}) \delta^{AD} \lambda_O^{BC}
\\ \nonumber && + ({3\over5} d_{NMO} + {i\over3} f_{NMO}) \delta^{BC} \lambda_O^{AD} - {2\over5} d_{NMO} \delta^{BD} \lambda_O^{AC}
\\ \nonumber && + ({\bf T}_{8\times10}^N)_{MP} \epsilon^{ABE} S_P^{CDE} + ({\bf T}_{8\times10}^N)^*_{MP} \epsilon^{CDE} S_P^{ABE}
\\ \nonumber && + ({\bf T}_{8\times27}^N)_{MU} S_U^{ABCD} \, ,
\end{eqnarray}
where ${\bf T}_{8\times10}^N$ and ${\bf T}_{8\times27}^N$ are the transition matrices satisfying $({\bf T}_{8\times10}^N)_{MP} = - ({\bf T}_{8\times10}^M)_{NP}$. We show their explicit forms in Appendix~\ref{sec:app}.

The chiral transformations of $\mathcal{T}^{(\mathbf{8}, \mathbf{1})}_{1\cdots5}$ are the same, while $\mathcal{T}^{(\mathbf{1}, \mathbf{8})}_6$ transform like their mirror fields:
\begin{eqnarray}
&& \left\{ \begin{array}{l}
\delta_5 \mathcal{T}^{(\mathbf{8}, \mathbf{1})}_{1\cdots5,M} = 0 \, ,
\\ \delta^{\vec a} \mathcal{T}^{(\mathbf{8}, \mathbf{1})}_{1\cdots5,M} = 2 a^N f_{NMO} \mathcal{T}^{(\mathbf{8}, \mathbf{1})}_{1\cdots5,O} \, ,
\\ \delta^{\vec b}_5 \mathcal{T}^{(\mathbf{8}, \mathbf{1})}_{1\cdots5,M} = 2 i b^N f_{NMO} \gamma_5 \mathcal{T}^{(\mathbf{8}, \mathbf{1})}_{1\cdots5,O} \, .
\end{array} \right .
\\ &&
\left\{ \begin{array}{l}
\delta_5 \mathcal{T}^{(\mathbf{1}, \mathbf{8})}_{6,M} = 0 \, ,
\\ \delta^{\vec a} \mathcal{T}^{(\mathbf{1}, \mathbf{8})}_{6,M} = 2 a^N f_{NMO} \mathcal{T}^{(\mathbf{1}, \mathbf{8})}_{6,O} \, ,
\\ \delta^{\vec b}_5 \mathcal{T}^{(\mathbf{1}, \mathbf{8})}_{6,M} = - 2 i b^N f_{NMO} \gamma_5 \mathcal{T}^{(\mathbf{1}, \mathbf{8})}_{6,O} \, .
\end{array} \right .
\end{eqnarray}
These chiral transformations properties are the same as those of $(\rho, a_1)$ belonging to the $[(\mathbf{8}, \mathbf{1})\oplus(\mathbf{1}, \mathbf{8})]$ chiral representation, or its mirror representation (Eqs.~(\ref{eq:meson81})).

\section{Tetraquark Currents of $J^{PC} = 1^{-+}$}
\label{sec:current}

In the previous sections we have studied the chiral (flavor) structure of tetraquarks. Those obtained flavor matrices and chiral transformation equations can be used to construct tetraquark currents and study their chiral transformation properties. We note that the chiral (flavor) structure of tetraquark currents is much more complicated than their color and Lorentz structures, and so the construction of tetraquark currents becomes easy (much easier) based on the studies of previous sections. As an example, in this section we show how to construct tetraquark currents of exotic quantum numbers $J^{PC} = 1^{-+}$, but belonging to the ``non-exotic'' chiral multiplet $[(\bar \mathbf{3}, \mathbf{3})\oplus(\mathbf{3}, \bar \mathbf{3})]$.

The tetraquark currents of $J^{PC} = 1^{-+}$ have been listed in Ref.~\cite{Chen:2008qw}, and so we just need to add the flavor matrices properly. The chiral multiplets $\mathcal{T}^{(\mathbf{3}, \bar\mathbf{3})}_1$ and $\mathcal{T}^{(\bar\mathbf{3},\mathbf{3})}_4$ both contain the diquark and antidiquark having the antisymmetry flavor structure $\mathbf{\bar 3_f}(qq) \otimes \mathbf{3_f}(\bar q \bar q)$, i.e., the flavor matrix $\epsilon^{ABE} \epsilon^{CDE}$. Inserting this matrix into Eqs.~(2) of Ref.~\cite{Chen:2008qv}, we obtain two independent flavor singlet tetraquark currents (the superscript $A$ is used to denote the antisymmetric flavor structure):
%
\begin{eqnarray}
\nonumber \mathcal{S}^A_{1\mu} &=& \epsilon^{ABE} \epsilon^{CDE} \big [ q_A^{aT} C \gamma_5 q_B^b (\bar{q}_C^a \gamma_\mu \gamma_5 C \bar{q}_D^{bT} - \bar{q}_C^b \gamma_\mu \gamma_5 C \bar{q}_D^{aT})
+ q_A^{aT} C \gamma_\mu \gamma_5 q_B^b (\bar{q}_C^a \gamma_5 C \bar{q}_D^{bT} - \bar{q}_C^b \gamma_5 C \bar{q}_D^{aT}) \big ] \, ,
\\ \mathcal{S}^A_{2\mu} &=& \epsilon^{ABE} \epsilon^{CDE} \big [ q_A^{aT} C \gamma^\nu q_B^b (\bar{q}_C^a \sigma_{\mu\nu} C \bar{q}_D^{bT} + \bar{q}_C^b \sigma_{\mu\nu} C \bar{q}_D^{aT})
+ q_A^{aT} C \sigma_{\mu\nu} q_B^b (\bar{q}_C^a \gamma^\nu C \bar{q}_D^{bT} + \bar{q}_C^b \gamma^\nu C \bar{q}_D^{aT}) \big ] \, ,
\label{eq:current11}
\end{eqnarray}
%
where the sum over repeated indices ($A,B,\cdots$ for flavor indices, $a, b, \cdots$ for color indices, and $\mu$, $\nu, \cdots$ for Dirac spinor indices) is taken. $C$ is the charge-conjugation matrix, $q_A^a$ and $q_B^b$ represent quarks, and $q_C^c$ and $q_D^d$ represent antiquarks. We can write them using the left- and right-handed quark fields $q_L = {1-\gamma_5 \over 2} q$ and  $q_R = {1+\gamma_5 \over 2} q$:
\begin{eqnarray}
\nonumber \mathcal{S}^A_{1\mu} &=& \epsilon^{ABE} \epsilon^{CDE} \big [ 2 q_{LA}^{aT} C q_{LB}^b (\bar{q}_{LC}^a \gamma_\mu C \bar{q}_{RD}^{bT} - \bar{q}_{LC}^b \gamma_\mu C \bar{q}_{RD}^{aT})
+ 2 q_{RA}^{aT} C q_{RB}^b (\bar{q}_{RA}^a \gamma_\mu C \bar{q}_{LB}^{bT} - \bar{q}_{RA}^b \gamma_\mu C \bar{q}_{LB}^{aT})
\\ && + 2 q_{LA}^{aT} C \gamma_\mu q_{RB}^b (\bar{q}_{LC}^a C \bar{q}_{LD}^{bT} - \bar{q}_{LC}^b C \bar{q}_{LD}^{aT})
+ 2 q_{RA}^{aT} C \gamma_\mu q_{LB}^b (\bar{q}_{RC}^a C \bar{q}_{RD}^{bT} - \bar{q}_{RC}^b C \bar{q}_{RD}^{aT}) \big ] \, ,
\\ \nonumber \mathcal{S}^A_{2\mu} &=& \epsilon^{ABE} \epsilon^{CDE} \big [ 2 q_{LA}^{aT} C \gamma^\nu q_{RB}^b (\bar{q}_{LC}^a \sigma_{\mu\nu} C \bar{q}_{LD}^{bT} + \bar{q}_{LC}^b \sigma_{\mu\nu} C \bar{q}_{LD}^{aT})
+ 2 q_{RA}^{aT} C \gamma^\nu q_{LB}^b (\bar{q}_{RC}^a \sigma_{\mu\nu} C \bar{q}_{RD}^{bT} + \bar{q}_{RC}^b \sigma_{\mu\nu} C \bar{q}_{RD}^{aT})
\\ \nonumber && + 2 q_{LA}^{aT} C \sigma_{\mu\nu} q_{LB}^b (\bar{q}_{LC}^a \gamma^\nu C \bar{q}_{RD}^{bT} + \bar{q}_{LC}^b \gamma^\nu C \bar{q}_{LD}^{aT})
+ 2 q_{RA}^{aT} C \sigma_{\mu\nu} q_{RB}^b (\bar{q}_{RC}^a \gamma^\nu C \bar{q}_{LD}^{bT} + \bar{q}_{RC}^b \gamma^\nu C \bar{q}_{LD}^{aT}) \big ] \, ,
\end{eqnarray}
from which we find that these two currents belong to both $\mathcal{T}^{(\mathbf{3}, \bar\mathbf{3})}_1$ and $\mathcal{T}^{(\bar\mathbf{3},\mathbf{3})}_4$. This is because these two chiral multiplets are related by the charge-conjugation transformation, and so if we want to construct tetraquark currents having a definite charge-conjugation parity, we need to use their combinations.

Inserting the flavor matrix $( \delta^{AC} \lambda_N^{DB} - \delta^{BC} \lambda_N^{DA} )$ into Eqs.~(2) of Ref.~\cite{Chen:2008qv}, we obtain two flavor octet tetraquark currents which also belong to both $\mathcal{T}^{(\mathbf{3}, \bar\mathbf{3})}_1$ and $\mathcal{T}^{(\bar\mathbf{3},\mathbf{3})}_4$:
%
\begin{eqnarray}
\nonumber \mathcal{O}^A_{1\mu,N} &=& ( \delta^{AC} \lambda_N^{DB} - \delta^{BC} \lambda_N^{DA} ) \big [ q_A^{aT} C \gamma_5 q_B^b (\bar{q}_C^a \gamma_\mu \gamma_5 C \bar{q}_D^{bT} - \bar{q}_C^b \gamma_\mu \gamma_5 C \bar{q}_D^{aT})
+ q_A^{aT} C \gamma_\mu \gamma_5 q_B^b (\bar{q}_C^a \gamma_5 C \bar{q}_D^{bT} - \bar{q}_C^b \gamma_5 C \bar{q}_D^{aT}) \big ] \, ,
\\ \mathcal{O}^A_{2\mu,N} &=& ( \delta^{AC} \lambda_N^{DB} - \delta^{BC} \lambda_N^{DA} ) \big [ q_A^{aT} C \gamma^\nu q_B^b (\bar{q}_C^a \sigma_{\mu\nu} C \bar{q}_D^{bT} + \bar{q}_C^b \sigma_{\mu\nu} C \bar{q}_D^{aT})
+ q_A^{aT} C \sigma_{\mu\nu} q_B^b (\bar{q}_C^a \gamma^\nu C \bar{q}_D^{bT} + \bar{q}_C^b \gamma^\nu C \bar{q}_D^{aT}) \big ] \, .
\end{eqnarray}
%
We can verify that $\mathcal{S}^A_{1\mu}$ and $\mathcal{O}^A_{1\mu,N}$ belong to the same chiral multiplet, while $\mathcal{S}^A_{2\mu}$ and $\mathcal{O}^A_{2\mu,N}$ belong to another chiral multiplet.

The chiral multiplet $\mathcal{T}^{(\mathbf{3}, \bar\mathbf{3})}_2$ and $\mathcal{T}^{(\bar\mathbf{3},\mathbf{3})}_5$ contains the diquark and the antidiquark having the symmetry flavor structure $\mathbf{6_f}(qq) \otimes \mathbf{\bar 6_f}(\bar q \bar q)$, i.e., the flavor matrix $( \delta^{AC} \delta^{BD} + \delta^{BC} \delta^{AD} )$, and they are also related by the charge-conjugation transformation. Inserting this matrix into Eqs.~(1) of Ref.~\cite{Chen:2008qv}, we obtain two independent flavor singlet tetraquark currents (the superscript $S$ is used to denote the symmetric flavor structure):
%
\begin{eqnarray}
\nonumber \mathcal{S}^S_{1\mu} &=& ( \delta^{AC} \delta^{BD} + \delta^{BC} \delta^{AD} ) \big [ q_A^{aT} C \gamma_5 q_B^b (\bar{q}_C^a \gamma_\mu \gamma_5 C \bar{q}_D^{bT} + \bar{q}_C^b \gamma_\mu
\gamma_5 C \bar{q}_D^{aT})
+ q_A^{aT} C \gamma_\mu \gamma_5 q_B^b (\bar{q}_C^a \gamma_5 C \bar{q}_D^{bT} + \bar{q}_C^b \gamma_5 C \bar{q}_D^{aT}) \big ] \, ,
\\ \mathcal{S}^S_{2\mu} &=& ( \delta^{AC} \delta^{BD} + \delta^{BC} \delta^{AD} ) \big [ q_A^{aT} C \gamma^\nu q_B^b (\bar{q}_C^a \sigma_{\mu\nu} C \bar{q}_D^{bT} - \bar{q}_C^b
\sigma_{\mu\nu} C \bar{q}_D^{aT})
+ q_A^{aT} C \sigma_{\mu\nu} q_B^b (\bar{q}_C^a \gamma^\nu C \bar{q}_D^{bT} - \bar{q}_C^b \gamma^\nu C \bar{q}_D^{aT}) \big ] \, .
\end{eqnarray}
%
They can be written as
%
\begin{eqnarray}
\nonumber \mathcal{S}^S_{1\mu} &=& \epsilon^{ABE} \epsilon^{CDE} \big [ 2 q_{LA}^{aT} C q_{LB}^b (\bar{q}_{LC}^a \gamma_\mu C \bar{q}_{RD}^{bT} + \bar{q}_{LC}^b \gamma_\mu C \bar{q}_{RD}^{aT})
+ 2 q_{RA}^{aT} C q_{RB}^b (\bar{q}_{RA}^a \gamma_\mu C \bar{q}_{LB}^{bT} + \bar{q}_{RA}^b \gamma_\mu C \bar{q}_{LB}^{aT})
\\ && + 2 q_{LA}^{aT} C \gamma_\mu q_{RB}^b (\bar{q}_{LC}^a C \bar{q}_{LD}^{bT} + \bar{q}_{LC}^b C \bar{q}_{LD}^{aT})
+ 2 q_{RA}^{aT} C \gamma_\mu q_{LB}^b (\bar{q}_{RC}^a C \bar{q}_{RD}^{bT} + \bar{q}_{RC}^b C \bar{q}_{RD}^{aT}) \big ] \, ,
\\ \nonumber \mathcal{S}^S_{2\mu} &=& \epsilon^{ABE} \epsilon^{CDE} \big [ 2 q_{LA}^{aT} C \gamma^\nu q_{RB}^b (\bar{q}_{LC}^a \sigma_{\mu\nu} C \bar{q}_{LD}^{bT} - \bar{q}_{LC}^b \sigma_{\mu\nu} C \bar{q}_{LD}^{aT})
+ 2 q_{RA}^{aT} C \gamma^\nu q_{LB}^b (\bar{q}_{RC}^a \sigma_{\mu\nu} C \bar{q}_{RD}^{bT} - \bar{q}_{RC}^b \sigma_{\mu\nu} C \bar{q}_{RD}^{aT})
\\ \nonumber && + 2 q_{LA}^{aT} C \sigma_{\mu\nu} q_{LB}^b (\bar{q}_{LC}^a \gamma^\nu C \bar{q}_{RD}^{bT} - \bar{q}_{LC}^b \gamma^\nu C \bar{q}_{LD}^{aT})
+ 2 q_{RA}^{aT} C \sigma_{\mu\nu} q_{RB}^b (\bar{q}_{RC}^a \gamma^\nu C \bar{q}_{LD}^{bT} - \bar{q}_{RC}^b \gamma^\nu C \bar{q}_{LD}^{aT}) \big ] \, ,
\end{eqnarray}
%
from which we find that these two currents belong to both $\mathcal{T}^{(\mathbf{3}, \bar\mathbf{3})}_2$ and $\mathcal{T}^{(\bar\mathbf{3},\mathbf{3})}_5$. 

Inserting the flavor matrix $( \delta^{AC} \lambda_N^{DB} + \delta^{BC} \lambda_N^{DA} )$ into Eqs.~(1) of Ref.~\cite{Chen:2008qv}, we obtain two flavor octet tetraquark currents which also belong to both $\mathcal{T}^{(\mathbf{3}, \bar\mathbf{3})}_2$ and $\mathcal{T}^{(\bar\mathbf{3},\mathbf{3})}_5$:
%
\begin{eqnarray}
\nonumber \mathcal{O}^S_{1\mu,N} &=& ( \delta^{AC} \lambda_N^{DB} + \delta^{BC} \lambda_N^{DA} ) \big [ q_A^{aT} C \gamma_5 q_B^b (\bar{q}_C^a \gamma_\mu \gamma_5 C \bar{q}_D^{bT} + \bar{q}_C^b \gamma_\mu
\gamma_5 C \bar{q}_D^{aT})
+ q_A^{aT} C \gamma_\mu \gamma_5 q_B^b (\bar{q}_C^a \gamma_5 C \bar{q}_D^{bT} + \bar{q}_C^b \gamma_5 C \bar{q}_D^{aT}) \big ] \, ,
\\ \mathcal{O}^S_{2\mu,N} &=& ( \delta^{AC} \lambda_N^{DB} + \delta^{BC} \lambda_N^{DA} ) \big [ q_A^{aT} C \gamma^\nu q_B^b (\bar{q}_C^a \sigma_{\mu\nu} C \bar{q}_D^{bT} - \bar{q}_C^b
\sigma_{\mu\nu} C \bar{q}_D^{aT})
+ q_A^{aT} C \sigma_{\mu\nu} q_B^b (\bar{q}_C^a \gamma^\nu C \bar{q}_D^{bT} - \bar{q}_C^b \gamma^\nu C \bar{q}_D^{aT}) \big ] \, .
\end{eqnarray}
%
We can verify that $\mathcal{S}^S_{1\mu}$ and $\mathcal{O}^S_{1\mu,N}$ belong to the same chiral multiplet, while $\mathcal{S}^S_{2\mu}$ and $\mathcal{O}^S_{2\mu,N}$ belong to another chiral multiplet.

The chiral multiplet $\mathcal{T}^{(\mathbf{3}, \bar\mathbf{3})}_3$ contains the diquark and the antidiquark having the antisymmetry flavor structure $\mathbf{\bar 3_f}(qq) \otimes \mathbf{3_f}(\bar q \bar q)$. However, after inserting the flavor matrix $\epsilon^{ABE} \epsilon^{CDE}$ into Eqs.~(2) of Ref.~\cite{Chen:2008qv}, we obtain the same tetraquark currents as Eqs.~(\ref{eq:current11}). We have verified that these two currents belonging to $\mathcal{T}^{(\mathbf{3}, \bar\mathbf{3})}_1$ and $\mathcal{T}^{(\bar\mathbf{3},\mathbf{3})}_4$. Therefore, there are no tetraquark currents belonging to the chiral multiplet $\mathcal{T}^{(\mathbf{3}, \bar\mathbf{3})}_3$, which have been proved to transform in an exotic way under chiral transformations (see Eqs.~(\ref{eq:tetraquark333}) and (\ref{eq:tetraquark338})).

The chiral transformation properties of tetraquark currents are different from those equations obtained in Sec.~\ref{sec:transformation} for the chiral space only, but their coefficients are similar. For example, the chiral transformation equations of $\mathcal{S}^A_{1\mu}$ are:
\begin{eqnarray}
\nonumber \delta_5 \mathcal{S}^A_{1\mu} &=& 2 i b^0 \times \epsilon^{ABE} \epsilon^{CDE} \big [ q_A^{aT} C q_B^b (\bar{q}_C^a \gamma_\mu \gamma_5 C \bar{q}_D^{bT} - \bar{q}_C^b \gamma_\mu \gamma_5 C \bar{q}_D^{aT})
+ q_A^{aT} C \gamma_\mu \gamma_5 q_B^b (\bar{q}_C^a C \bar{q}_D^{bT} - \bar{q}_C^b C \bar{q}_D^{aT}) \big ]
\\ \nonumber &=& 2 i b^0 \mathcal{S}^{\prime A}_{1\mu} \, ,
\\ \delta^{\vec a} \mathcal{S}^A_{1\mu} &=& 0 \, ,
\label{eq:current331}
\\ \nonumber \delta^{\vec b}_5 \mathcal{S}^A_{1\mu} &=& 2 i b^N \times ( \delta^{AC} \lambda_N^{DB} - \delta^{BC} \lambda_N^{DA} ) \big [ q_A^{aT} C q_B^b (\bar{q}_C^a \gamma_\mu \gamma_5 C \bar{q}_D^{bT} - \bar{q}_C^b \gamma_\mu \gamma_5 C \bar{q}_D^{aT})
+ q_A^{aT} C \gamma_\mu \gamma_5 q_B^b (\bar{q}_C^a C \bar{q}_D^{bT} - \bar{q}_C^b C \bar{q}_D^{aT}) \big ] \, ,
\\ \nonumber &=& 2 i b^N \mathcal{O}^{\prime A}_{1\mu,N} \, ,
\end{eqnarray}
where $\mathcal{S}^{\prime A}_{1\mu}$ and $\mathcal{O}^{\prime A}_{1\mu,N}$ are the tetraquark currents having quantum numbers $J^{PC} = 1^{++}$ and belonging to the chiral representation $[(\mathbf{3},\bar\mathbf{3})\oplus(\bar\mathbf{3},\mathbf{3})]$ (indeed they also belong to the chiral multiplets $\mathcal{T}^{(\mathbf{3}, \bar\mathbf{3})}_1$ and $\mathcal{T}^{(\bar\mathbf{3},\mathbf{3})}_4$). Therefore, we have shown that the formulae obtained in Sec.~\ref{sec:transformation} can be easily applied to calculate the chiral transformation equations of the relevant tetraquark currents. Moreover, since we have verified that there are no local tetraquark currents of $J^{PC} = 1^{-+}$ belonging to the chiral multiplet $\mathcal{T}^{(\mathbf{3}, \bar\mathbf{3})}_3$, we arrive at a conclusion that all the local tetraquark currents of $J^{PC} = 1^{-+}$ and $[(\mathbf{3},\bar\mathbf{3})\oplus(\bar\mathbf{3},\mathbf{3})]$ transform in the way similarly to the relevant quark-antiquark mesons of $[(\mathbf{3},\bar\mathbf{3})\oplus(\bar\mathbf{3},\mathbf{3})]$.

Since in this paper we concentrate on the chiral (flavor) structure of tetraquarks, we do not discuss tetraquark currents any more. We just note that these tetraquark currents can be used in the methods of Lattice QCD and QCD sum rule, and we shall also use them in our future QCD sum rule studies.

\section{Conclusion and Summary}
\label{sec:summary}

We have systematically investigated the chiral (flavor) structure of tetraquarks, and found all the chiral multiplets. We only consider the chiral (flavor) structure and other degrees of freedom remain undetermined. Then we concentrate on the tetraquarks belonging to the ``non-exotic'' $[(\bar \mathbf{3}, \mathbf{3})\oplus(\mathbf{3}, \bar \mathbf{3})]$ and $[(\mathbf{8}, \mathbf{1})\oplus(\mathbf{1}, \mathbf{8})]$ chiral multiplets as well as their mirror multiplets, which have the same representations as the lowest level $\bar q q$ chiral multiplets. We have studied their behaviors under the $U(1)_V$, $U(1)_A$, $SU(3)_V$ and $SU(3)_A$ chiral transformations. We find that most of them contain one pair of quark and antiquark which have the same chirality and can be combined to be a chiral singlet, and so they can be constructed by adding one chiral singlet quark-antiquark pair to the lowest level $\bar q q$ chiral multiplets. Consequently, under chiral transformations they transform exactly like the lowest level $[(\bar \mathbf{3}, \mathbf{3})\oplus(\mathbf{3}, \bar \mathbf{3})]$ chiral multiplet $(\sigma,\pi)$ and the $[(\mathbf{8}, \mathbf{1})\oplus(\mathbf{1}, \mathbf{8})]$ chiral multiplet $(\rho, a_1)$. There is only one exception, $\mathcal{T}^{(\bar\mathbf{3},\mathbf{3})}_{3}$, whose quark-antiquark pairs all have the opposite chirality, and it transforms differently from other $[(\bar \mathbf{3}, \mathbf{3})\oplus(\mathbf{3}, \bar \mathbf{3})]$ chiral multiplets.

The flavor matrices and chiral transformation equations obtained in the chiral (flavor) space of tetraquarks can be used to construct tetraquark currents and study their chiral transformation properties. As an example, we construct tetraquark currents of exotic quantum numbers $J^{PC} = 1^{-+}$, but belonging to the ``non-exotic'' chiral multiplet $[(\bar \mathbf{3}, \mathbf{3})\oplus(\mathbf{3}, \bar \mathbf{3})]$. We also calculate the chiral transformation equations of one tetraquark current $\mathcal{S}^A_{1\mu}$, and find the coefficients of these equations (Eqs.~(\ref{eq:current331})) are similar to those for tetraquarks belonging to the same chiral multiplet $[(\bar \mathbf{3}, \mathbf{3})\oplus(\mathbf{3}, \bar \mathbf{3})]$, but obtained in the chiral space only (Eqs.~(\ref{eq:tetraquark331})).

Since there are always multi-quark components in the Fock space expansion of physical hadron states, it is worth studying what remains unchanged in this expansion. In this paper we have shown that as long as the tetraquark and the $\bar q q$ meson belong to the same chiral representation and the tetraquark contains one quark and one antiquark having the same chirality ($\bar q_L q_L + \bar q_R q_R$), they transform in the same way under chiral transformations. We find that among the five $[(\bar \mathbf{3}, \mathbf{3})\oplus(\mathbf{3}, \bar \mathbf{3})]$ and six $[(\mathbf{8}, \mathbf{1})\oplus(\mathbf{1}, \mathbf{8})]$ tetraquark chiral multiplets shown in Tabs~\ref{tab:singlet} and \ref{tab:octet}, ten of them satisfy this condition.

This study is for the tetraquark case, the pentaquark case is also interesting, which we shall study in the future. We shall also perform chiral transformations on other exotic tetraquarks in order to make a complete analysis, which is more complicated and takes more time.

\section*{Acknowledgments}
\label{ack}

This work is partly supported by the National Natural Science Foundation of China under Grant No. 11147140, and the Scientific Research Foundation for the Returned Overseas Chinese Scholars, State Education Ministry.

\appendix

\section{Transition Matrices}
\label{sec:app}

Since it takes much more space to show the transition matrices ${\bf T}_{8\times10}^N$, we show the conjugate-transpose ones ${\bf T}_{8\times10}^{N\dagger}$ here
\begin{eqnarray}
\nonumber {\bf T}_{8\times10}^{1\dagger} &=& \left(
\begin{array}{cccccccc}
 0 & 0 & 0 & \frac{1}{2} & -\frac{i}{2} & 0 & 0 & 0 \\
 0 & 0 & 0 & 0 & 0 & \frac{1}{2 \sqrt{3}} & -\frac{i}{2 \sqrt{3}} & 0 \\
 0 & 0 & 0 & -\frac{1}{2 \sqrt{3}} & \frac{i}{2 \sqrt{3}} & 0 & 0 & 0 \\
 0 & 0 & 0 & 0 & 0 & -\frac{1}{2} & \frac{i}{2} & 0 \\
 0 & 0 & -\frac{1}{2 \sqrt{3}} & 0 & 0 & 0 & 0 & -\frac{1}{2} \\
 0 & -\frac{i}{\sqrt{6}} & 0 & 0 & 0 & 0 & 0 & 0 \\
 0 & 0 & -\frac{1}{2 \sqrt{3}} & 0 & 0 & 0 & 0 & \frac{1}{2} \\
 0 & 0 & 0 & -\frac{1}{2 \sqrt{3}} & -\frac{i}{2 \sqrt{3}} & 0 & 0 & 0 \\
 0 & 0 & 0 & 0 & 0 & \frac{1}{2 \sqrt{3}} & \frac{i}{2 \sqrt{3}} & 0 \\
 0 & 0 & 0 & 0 & 0 & 0 & 0 & 0
\end{array}
\right) \, ,
\\ \nonumber {\bf T}_{8\times10}^{2\dagger} &=& \left(
\begin{array}{cccccccc}
 0 & 0 & 0 & -\frac{i}{2} & -\frac{1}{2} & 0 & 0 & 0 \\
 0 & 0 & 0 & 0 & 0 & -\frac{i}{2 \sqrt{3}} & -\frac{1}{2 \sqrt{3}} & 0 \\
 0 & 0 & 0 & -\frac{i}{2 \sqrt{3}} & -\frac{1}{2 \sqrt{3}} & 0 & 0 & 0 \\
 0 & 0 & 0 & 0 & 0 & -\frac{i}{2} & -\frac{1}{2} & 0 \\
 0 & 0 & \frac{i}{2 \sqrt{3}} & 0 & 0 & 0 & 0 & \frac{i}{2} \\
 \frac{i}{\sqrt{6}} & 0 & 0 & 0 & 0 & 0 & 0 & 0 \\
 0 & 0 & -\frac{i}{2 \sqrt{3}} & 0 & 0 & 0 & 0 & \frac{i}{2} \\
 0 & 0 & 0 & \frac{i}{2 \sqrt{3}} & -\frac{1}{2 \sqrt{3}} & 0 & 0 & 0 \\
 0 & 0 & 0 & 0 & 0 & \frac{i}{2 \sqrt{3}} & -\frac{1}{2 \sqrt{3}} & 0 \\
 0 & 0 & 0 & 0 & 0 & 0 & 0 & 0
\end{array}
\right) \, ,
\\ \nonumber {\bf T}_{8\times10}^{3\dagger} &=& \left(
\begin{array}{cccccccc}
 0 & 0 & 0 & 0 & 0 & 0 & 0 & 0 \\
 0 & 0 & 0 & -\frac{1}{\sqrt{3}} & \frac{i}{\sqrt{3}} & 0 & 0 & 0 \\
 0 & 0 & 0 & 0 & 0 & -\frac{1}{\sqrt{3}} & \frac{i}{\sqrt{3}} & 0 \\
 0 & 0 & 0 & 0 & 0 & 0 & 0 & 0 \\
 \frac{1}{2 \sqrt{3}} & -\frac{i}{2 \sqrt{3}} & 0 & 0 & 0 & 0 & 0 & 0 \\
 0 & 0 & 0 & 0 & 0 & 0 & 0 & \frac{1}{\sqrt{2}} \\
 \frac{1}{2 \sqrt{3}} & \frac{i}{2 \sqrt{3}} & 0 & 0 & 0 & 0 & 0 & 0 \\
 0 & 0 & 0 & 0 & 0 & \frac{1}{2 \sqrt{3}} & \frac{i}{2 \sqrt{3}} & 0 \\
 0 & 0 & 0 & \frac{1}{2 \sqrt{3}} & \frac{i}{2 \sqrt{3}} & 0 & 0 & 0 \\
 0 & 0 & 0 & 0 & 0 & 0 & 0 & 0
\end{array}
\right) \, ,
\\ \nonumber {\bf T}_{8\times10}^{4\dagger} &=& \left(
\begin{array}{cccccccc}
 -\frac{1}{2} & \frac{i}{2} & 0 & 0 & 0 & 0 & 0 & 0 \\
 0 & 0 & \frac{1}{\sqrt{3}} & 0 & 0 & 0 & 0 & 0 \\
 \frac{1}{2 \sqrt{3}} & \frac{i}{2 \sqrt{3}} & 0 & 0 & 0 & 0 & 0 & 0 \\
 0 & 0 & 0 & 0 & 0 & 0 & 0 & 0 \\
 0 & 0 & 0 & 0 & 0 & -\frac{1}{2 \sqrt{3}} & -\frac{i}{2 \sqrt{3}} & 0 \\
 0 & 0 & 0 & 0 & \frac{i}{\sqrt{6}} & 0 & 0 & 0 \\
 0 & 0 & 0 & 0 & 0 & -\frac{1}{2 \sqrt{3}} & \frac{i}{2 \sqrt{3}} & 0 \\
 \frac{1}{2 \sqrt{3}} & -\frac{i}{2 \sqrt{3}} & 0 & 0 & 0 & 0 & 0 & 0 \\
 0 & 0 & -\frac{1}{2 \sqrt{3}} & 0 & 0 & 0 & 0 & \frac{1}{2} \\
 0 & 0 & 0 & 0 & 0 & \frac{1}{2} & \frac{i}{2} & 0
\end{array}
\right) \, ,
\\ \nonumber {\bf T}_{8\times10}^{5\dagger} &=& \left(
\begin{array}{cccccccc}
 \frac{i}{2} & \frac{1}{2} & 0 & 0 & 0 & 0 & 0 & 0 \\
 0 & 0 & -\frac{i}{\sqrt{3}} & 0 & 0 & 0 & 0 & 0 \\
 -\frac{i}{2 \sqrt{3}} & \frac{1}{2 \sqrt{3}} & 0 & 0 & 0 & 0 & 0 & 0 \\
 0 & 0 & 0 & 0 & 0 & 0 & 0 & 0 \\
 0 & 0 & 0 & 0 & 0 & \frac{i}{2 \sqrt{3}} & -\frac{1}{2 \sqrt{3}} & 0 \\
 0 & 0 & 0 & -\frac{i}{\sqrt{6}} & 0 & 0 & 0 & 0 \\
 0 & 0 & 0 & 0 & 0 & -\frac{i}{2 \sqrt{3}} & -\frac{1}{2 \sqrt{3}} & 0 \\
 \frac{i}{2 \sqrt{3}} & \frac{1}{2 \sqrt{3}} & 0 & 0 & 0 & 0 & 0 & 0 \\
 0 & 0 & -\frac{i}{2 \sqrt{3}} & 0 & 0 & 0 & 0 & \frac{i}{2} \\
 0 & 0 & 0 & 0 & 0 & \frac{i}{2} & -\frac{1}{2} & 0
\end{array}
\right) \, ,
\\ \nonumber {\bf T}_{8\times10}^{6\dagger} &=& \left(
\begin{array}{cccccccc}
 0 & 0 & 0 & 0 & 0 & 0 & 0 & 0 \\
 -\frac{1}{2 \sqrt{3}} & \frac{i}{2 \sqrt{3}} & 0 & 0 & 0 & 0 & 0 & 0 \\
 0 & 0 & \frac{1}{\sqrt{3}} & 0 & 0 & 0 & 0 & 0 \\
 \frac{1}{2} & \frac{i}{2} & 0 & 0 & 0 & 0 & 0 & 0 \\
 0 & 0 & 0 & \frac{1}{2 \sqrt{3}} & -\frac{i}{2 \sqrt{3}} & 0 & 0 & 0 \\
 0 & 0 & 0 & 0 & 0 & 0 & -\frac{i}{\sqrt{6}} & 0 \\
 0 & 0 & 0 & \frac{1}{2 \sqrt{3}} & \frac{i}{2 \sqrt{3}} & 0 & 0 & 0 \\
 0 & 0 & -\frac{1}{2 \sqrt{3}} & 0 & 0 & 0 & 0 & -\frac{1}{2} \\
 -\frac{1}{2 \sqrt{3}} & -\frac{i}{2 \sqrt{3}} & 0 & 0 & 0 & 0 & 0 & 0 \\
 0 & 0 & 0 & -\frac{1}{2} & -\frac{i}{2} & 0 & 0 & 0
\end{array}
\right) \, ,
\\ \nonumber {\bf T}_{8\times10}^{7\dagger} &=& \left(
\begin{array}{cccccccc}
 0 & 0 & 0 & 0 & 0 & 0 & 0 & 0 \\
 \frac{i}{2 \sqrt{3}} & \frac{1}{2 \sqrt{3}} & 0 & 0 & 0 & 0 & 0 & 0 \\
 0 & 0 & -\frac{i}{\sqrt{3}} & 0 & 0 & 0 & 0 & 0 \\
 -\frac{i}{2} & \frac{1}{2} & 0 & 0 & 0 & 0 & 0 & 0 \\
 0 & 0 & 0 & \frac{i}{2 \sqrt{3}} & \frac{1}{2 \sqrt{3}} & 0 & 0 & 0 \\
 0 & 0 & 0 & 0 & 0 & \frac{i}{\sqrt{6}} & 0 & 0 \\
 0 & 0 & 0 & -\frac{i}{2 \sqrt{3}} & \frac{1}{2 \sqrt{3}} & 0 & 0 & 0 \\
 0 & 0 & -\frac{i}{2 \sqrt{3}} & 0 & 0 & 0 & 0 & -\frac{i}{2} \\
 -\frac{i}{2 \sqrt{3}} & \frac{1}{2 \sqrt{3}} & 0 & 0 & 0 & 0 & 0 & 0 \\
 0 & 0 & 0 & -\frac{i}{2} & \frac{1}{2} & 0 & 0 & 0
\end{array}
\right) \, ,
\\ \nonumber {\bf T}_{8\times10}^{8\dagger} &=& \left(
\begin{array}{cccccccc}
 0 & 0 & 0 & 0 & 0 & 0 & 0 & 0 \\
 0 & 0 & 0 & 0 & 0 & 0 & 0 & 0 \\
 0 & 0 & 0 & 0 & 0 & 0 & 0 & 0 \\
 0 & 0 & 0 & 0 & 0 & 0 & 0 & 0 \\
 \frac{1}{2} & -\frac{i}{2} & 0 & 0 & 0 & 0 & 0 & 0 \\
 0 & 0 & -\frac{1}{\sqrt{2}} & 0 & 0 & 0 & 0 & 0 \\
 -\frac{1}{2} & -\frac{i}{2} & 0 & 0 & 0 & 0 & 0 & 0 \\
 0 & 0 & 0 & 0 & 0 & \frac{1}{2} & \frac{i}{2} & 0 \\
 0 & 0 & 0 & -\frac{1}{2} & -\frac{i}{2} & 0 & 0 & 0 \\
 0 & 0 & 0 & 0 & 0 & 0 & 0 & 0
\end{array}
\right) \, .
\end{eqnarray}
Since it takes much more space to show the transition matrices ${\bf T}_{8\times27}^N$, we show the conjugate-transpose ones ${\bf T}_{8\times27}^{N\dagger}$ here
\begin{eqnarray}
\nonumber {\bf T}_{8\times27}^{1\dagger} &=& \left(
\begin{array}{cccccccc}
 \frac{1}{\sqrt{30}} & 0 & 0 & 0 & 0 & 0 & 0 & 0 \\
 0 & 0 & 0 & 0 & 0 & 0 & 0 & \sqrt{\frac{3}{5}} \\
 0 & 0 & 0 & 0 & 0 & 0 & 0 & 0 \\
 0 & 0 & 0 & 0 & 0 & 0 & 0 & \sqrt{\frac{3}{5}} \\
 0 & 0 & 0 & 0 & 0 & \frac{1}{\sqrt{30}} & \frac{i}{\sqrt{30}} & 0 \\
 0 & 0 & 0 & \frac{1}{\sqrt{30}} & \frac{i}{\sqrt{30}} & 0 & 0 & 0 \\
 0 & 0 & 0 & \frac{1}{\sqrt{30}} & -\frac{i}{\sqrt{30}} & 0 & 0 & 0 \\
 0 & 0 & 0 & 0 & 0 & \frac{1}{\sqrt{30}} & -\frac{i}{\sqrt{30}} & 0 \\
 1 & i & 0 & 0 & 0 & 0 & 0 & 0 \\
 0 & 0 & 1 & 0 & 0 & 0 & 0 & 0 \\
 -\sqrt{\frac{2}{3}} & 0 & 0 & 0 & 0 & 0 & 0 & 0 \\
 0 & 0 & 1 & 0 & 0 & 0 & 0 & 0 \\
 1 & -i & 0 & 0 & 0 & 0 & 0 & 0 \\
 0 & 0 & 0 & \frac{1}{\sqrt{2}} & \frac{i}{\sqrt{2}} & 0 & 0 & 0 \\
 0 & 0 & 0 & 0 & 0 & -\frac{1}{\sqrt{6}} & -\frac{i}{\sqrt{6}} & 0 \\
 0 & 0 & 0 & \frac{1}{\sqrt{6}} & \frac{i}{\sqrt{6}} & 0 & 0 & 0 \\
 0 & 0 & 0 & 0 & 0 & \frac{1}{\sqrt{2}} & \frac{i}{\sqrt{2}} & 0 \\
 0 & 0 & 0 & 0 & 0 & \frac{1}{\sqrt{2}} & -\frac{i}{\sqrt{2}} & 0 \\
 0 & 0 & 0 & \frac{1}{\sqrt{6}} & -\frac{i}{\sqrt{6}} & 0 & 0 & 0 \\
 0 & 0 & 0 & 0 & 0 & -\frac{1}{\sqrt{6}} & \frac{i}{\sqrt{6}} & 0 \\
 0 & 0 & 0 & \frac{1}{\sqrt{2}} & -\frac{i}{\sqrt{2}} & 0 & 0 & 0 \\
 0 & 0 & 0 & 0 & 0 & 0 & 0 & 0 \\
 0 & 0 & 0 & 0 & 0 & 0 & 0 & 0 \\
 0 & 0 & 0 & 0 & 0 & 0 & 0 & 0 \\
 0 & 0 & 0 & 0 & 0 & 0 & 0 & 0 \\
 0 & 0 & 0 & 0 & 0 & 0 & 0 & 0 \\
 0 & 0 & 0 & 0 & 0 & 0 & 0 & 0
\end{array}
\right) \, ,
\\ \nonumber {\bf T}_{8\times27}^{2\dagger} &=& \left(
\begin{array}{cccccccc}
 0 & \frac{1}{\sqrt{30}} & 0 & 0 & 0 & 0 & 0 & 0 \\
 0 & 0 & 0 & 0 & 0 & 0 & 0 & i \sqrt{\frac{3}{5}} \\
 0 & 0 & 0 & 0 & 0 & 0 & 0 & 0 \\
 0 & 0 & 0 & 0 & 0 & 0 & 0 & -i \sqrt{\frac{3}{5}} \\
 0 & 0 & 0 & 0 & 0 & \frac{i}{\sqrt{30}} & -\frac{1}{\sqrt{30}} & 0 \\
 0 & 0 & 0 & -\frac{i}{\sqrt{30}} & \frac{1}{\sqrt{30}} & 0 & 0 & 0 \\
 0 & 0 & 0 & \frac{i}{\sqrt{30}} & \frac{1}{\sqrt{30}} & 0 & 0 & 0 \\
 0 & 0 & 0 & 0 & 0 & -\frac{i}{\sqrt{30}} & -\frac{1}{\sqrt{30}} & 0 \\
 i & -1 & 0 & 0 & 0 & 0 & 0 & 0 \\
 0 & 0 & i & 0 & 0 & 0 & 0 & 0 \\
 0 & -\sqrt{\frac{2}{3}} & 0 & 0 & 0 & 0 & 0 & 0 \\
 0 & 0 & -i & 0 & 0 & 0 & 0 & 0 \\
 -i & -1 & 0 & 0 & 0 & 0 & 0 & 0 \\
 0 & 0 & 0 & \frac{i}{\sqrt{2}} & -\frac{1}{\sqrt{2}} & 0 & 0 & 0 \\
 0 & 0 & 0 & 0 & 0 & -\frac{i}{\sqrt{6}} & \frac{1}{\sqrt{6}} & 0 \\
 0 & 0 & 0 & -\frac{i}{\sqrt{6}} & \frac{1}{\sqrt{6}} & 0 & 0 & 0 \\
 0 & 0 & 0 & 0 & 0 & -\frac{i}{\sqrt{2}} & \frac{1}{\sqrt{2}} & 0 \\
 0 & 0 & 0 & 0 & 0 & \frac{i}{\sqrt{2}} & \frac{1}{\sqrt{2}} & 0 \\
 0 & 0 & 0 & \frac{i}{\sqrt{6}} & \frac{1}{\sqrt{6}} & 0 & 0 & 0 \\
 0 & 0 & 0 & 0 & 0 & \frac{i}{\sqrt{6}} & \frac{1}{\sqrt{6}} & 0 \\
 0 & 0 & 0 & -\frac{i}{\sqrt{2}} & -\frac{1}{\sqrt{2}} & 0 & 0 & 0 \\
 0 & 0 & 0 & 0 & 0 & 0 & 0 & 0 \\
 0 & 0 & 0 & 0 & 0 & 0 & 0 & 0 \\
 0 & 0 & 0 & 0 & 0 & 0 & 0 & 0 \\
 0 & 0 & 0 & 0 & 0 & 0 & 0 & 0 \\
 0 & 0 & 0 & 0 & 0 & 0 & 0 & 0 \\
 0 & 0 & 0 & 0 & 0 & 0 & 0 & 0
\end{array}
\right) \, ,
\\ \nonumber {\bf T}_{8\times27}^{3\dagger} &=& \left(
\begin{array}{cccccccc}
 0 & 0 & \frac{1}{\sqrt{30}} & 0 & 0 & 0 & 0 & 0 \\
 0 & 0 & 0 & 0 & 0 & 0 & 0 & 0 \\
 0 & 0 & 0 & 0 & 0 & 0 & 0 & \sqrt{\frac{6}{5}} \\
 0 & 0 & 0 & 0 & 0 & 0 & 0 & 0 \\
 0 & 0 & 0 & \frac{1}{\sqrt{30}} & \frac{i}{\sqrt{30}} & 0 & 0 & 0 \\
 0 & 0 & 0 & 0 & 0 & -\frac{1}{\sqrt{30}} & -\frac{i}{\sqrt{30}} & 0 \\
 0 & 0 & 0 & 0 & 0 & -\frac{1}{\sqrt{30}} & \frac{i}{\sqrt{30}} & 0 \\
 0 & 0 & 0 & \frac{1}{\sqrt{30}} & -\frac{i}{\sqrt{30}} & 0 & 0 & 0 \\
 0 & 0 & 0 & 0 & 0 & 0 & 0 & 0 \\
 1 & i & 0 & 0 & 0 & 0 & 0 & 0 \\
 0 & 0 & 2 \sqrt{\frac{2}{3}} & 0 & 0 & 0 & 0 & 0 \\
 1 & -i & 0 & 0 & 0 & 0 & 0 & 0 \\
 0 & 0 & 0 & 0 & 0 & 0 & 0 & 0 \\
 0 & 0 & 0 & 0 & 0 & 0 & 0 & 0 \\
 0 & 0 & 0 & \sqrt{\frac{2}{3}} & i \sqrt{\frac{2}{3}} & 0 & 0 & 0 \\
 0 & 0 & 0 & 0 & 0 & \sqrt{\frac{2}{3}} & i \sqrt{\frac{2}{3}} & 0 \\
 0 & 0 & 0 & 0 & 0 & 0 & 0 & 0 \\
 0 & 0 & 0 & 0 & 0 & 0 & 0 & 0 \\
 0 & 0 & 0 & 0 & 0 & \sqrt{\frac{2}{3}} & -i \sqrt{\frac{2}{3}} & 0 \\
 0 & 0 & 0 & \sqrt{\frac{2}{3}} & -i \sqrt{\frac{2}{3}} & 0 & 0 & 0 \\
 0 & 0 & 0 & 0 & 0 & 0 & 0 & 0 \\
 0 & 0 & 0 & 0 & 0 & 0 & 0 & 0 \\
 0 & 0 & 0 & 0 & 0 & 0 & 0 & 0 \\
 0 & 0 & 0 & 0 & 0 & 0 & 0 & 0 \\
 0 & 0 & 0 & 0 & 0 & 0 & 0 & 0 \\
 0 & 0 & 0 & 0 & 0 & 0 & 0 & 0 \\
 0 & 0 & 0 & 0 & 0 & 0 & 0 & 0
\end{array}
\right) \, ,
\\ \nonumber {\bf T}_{8\times27}^{4\dagger} &=& \left(
\begin{array}{cccccccc}
 0 & 0 & 0 & -\sqrt{\frac{3}{10}} & 0 & 0 & 0 & 0 \\
 0 & 0 & 0 & 0 & 0 & -\frac{1}{\sqrt{5}} & \frac{i}{\sqrt{5}} & 0 \\
 0 & 0 & 0 & -\sqrt{\frac{2}{5}} & 0 & 0 & 0 & 0 \\
 0 & 0 & 0 & 0 & 0 & -\frac{1}{\sqrt{5}} & -\frac{i}{\sqrt{5}} & 0 \\
 0 & 0 & \frac{1}{\sqrt{30}} & 0 & 0 & 0 & 0 & \frac{3}{\sqrt{10}} \\
 \frac{1}{\sqrt{30}} & -\frac{i}{\sqrt{30}} & 0 & 0 & 0 & 0 & 0 & 0 \\
 \frac{1}{\sqrt{30}} & \frac{i}{\sqrt{30}} & 0 & 0 & 0 & 0 & 0 & 0 \\
 0 & 0 & \frac{1}{\sqrt{30}} & 0 & 0 & 0 & 0 & \frac{3}{\sqrt{10}} \\
 0 & 0 & 0 & 0 & 0 & 0 & 0 & 0 \\
 0 & 0 & 0 & 0 & 0 & 0 & 0 & 0 \\
 0 & 0 & 0 & 0 & 0 & 0 & 0 & 0 \\
 0 & 0 & 0 & 0 & 0 & 0 & 0 & 0 \\
 0 & 0 & 0 & 0 & 0 & 0 & 0 & 0 \\
 \frac{1}{\sqrt{2}} & \frac{i}{\sqrt{2}} & 0 & 0 & 0 & 0 & 0 & 0 \\
 0 & 0 & \sqrt{\frac{2}{3}} & 0 & 0 & 0 & 0 & 0 \\
 \frac{1}{\sqrt{6}} & -\frac{i}{\sqrt{6}} & 0 & 0 & 0 & 0 & 0 & 0 \\
 0 & 0 & 0 & 0 & 0 & 0 & 0 & 0 \\
 0 & 0 & 0 & 0 & 0 & 0 & 0 & 0 \\
 \frac{1}{\sqrt{6}} & \frac{i}{\sqrt{6}} & 0 & 0 & 0 & 0 & 0 & 0 \\
 0 & 0 & \sqrt{\frac{2}{3}} & 0 & 0 & 0 & 0 & 0 \\
 \frac{1}{\sqrt{2}} & -\frac{i}{\sqrt{2}} & 0 & 0 & 0 & 0 & 0 & 0 \\
 0 & 0 & 0 & 1 & i & 0 & 0 & 0 \\
 0 & 0 & 0 & 0 & 0 & \frac{1}{\sqrt{2}} & \frac{i}{\sqrt{2}} & 0 \\
 0 & 0 & 0 & 0 & 0 & 0 & 0 & 0 \\
 0 & 0 & 0 & 0 & 0 & 0 & 0 & 0 \\
 0 & 0 & 0 & 0 & 0 & \frac{1}{\sqrt{2}} & -\frac{i}{\sqrt{2}} & 0 \\
 0 & 0 & 0 & 1 & -i & 0 & 0 & 0
\end{array}
\right) \, ,
\\ \nonumber {\bf T}_{8\times27}^{5\dagger} &=& \left(
\begin{array}{cccccccc}
 0 & 0 & 0 & 0 & -\sqrt{\frac{3}{10}} & 0 & 0 & 0 \\
 0 & 0 & 0 & 0 & 0 & -\frac{i}{\sqrt{5}} & -\frac{1}{\sqrt{5}} & 0 \\
 0 & 0 & 0 & 0 & -\sqrt{\frac{2}{5}} & 0 & 0 & 0 \\
 0 & 0 & 0 & 0 & 0 & \frac{i}{\sqrt{5}} & -\frac{1}{\sqrt{5}} & 0 \\
 0 & 0 & \frac{i}{\sqrt{30}} & 0 & 0 & 0 & 0 & \frac{3 i}{\sqrt{10}} \\
 \frac{i}{\sqrt{30}} & \frac{1}{\sqrt{30}} & 0 & 0 & 0 & 0 & 0 & 0 \\
 -\frac{i}{\sqrt{30}} & \frac{1}{\sqrt{30}} & 0 & 0 & 0 & 0 & 0 & 0 \\
 0 & 0 & -\frac{i}{\sqrt{30}} & 0 & 0 & 0 & 0 & -\frac{3 i}{\sqrt{10}} \\
 0 & 0 & 0 & 0 & 0 & 0 & 0 & 0 \\
 0 & 0 & 0 & 0 & 0 & 0 & 0 & 0 \\
 0 & 0 & 0 & 0 & 0 & 0 & 0 & 0 \\
 0 & 0 & 0 & 0 & 0 & 0 & 0 & 0 \\
 0 & 0 & 0 & 0 & 0 & 0 & 0 & 0 \\
 \frac{i}{\sqrt{2}} & -\frac{1}{\sqrt{2}} & 0 & 0 & 0 & 0 & 0 & 0 \\
 0 & 0 & i \sqrt{\frac{2}{3}} & 0 & 0 & 0 & 0 & 0 \\
 \frac{i}{\sqrt{6}} & \frac{1}{\sqrt{6}} & 0 & 0 & 0 & 0 & 0 & 0 \\
 0 & 0 & 0 & 0 & 0 & 0 & 0 & 0 \\
 0 & 0 & 0 & 0 & 0 & 0 & 0 & 0 \\
 -\frac{i}{\sqrt{6}} & \frac{1}{\sqrt{6}} & 0 & 0 & 0 & 0 & 0 & 0 \\
 0 & 0 & -i \sqrt{\frac{2}{3}} & 0 & 0 & 0 & 0 & 0 \\
 -\frac{i}{\sqrt{2}} & -\frac{1}{\sqrt{2}} & 0 & 0 & 0 & 0 & 0 & 0 \\
 0 & 0 & 0 & i & -1 & 0 & 0 & 0 \\
 0 & 0 & 0 & 0 & 0 & \frac{i}{\sqrt{2}} & -\frac{1}{\sqrt{2}} & 0 \\
 0 & 0 & 0 & 0 & 0 & 0 & 0 & 0 \\
 0 & 0 & 0 & 0 & 0 & 0 & 0 & 0 \\
 0 & 0 & 0 & 0 & 0 & -\frac{i}{\sqrt{2}} & -\frac{1}{\sqrt{2}} & 0 \\
 0 & 0 & 0 & -i & -1 & 0 & 0 & 0
\end{array}
\right) \, ,
\\ \nonumber {\bf T}_{8\times27}^{6\dagger} &=& \left(
\begin{array}{cccccccc}
 0 & 0 & 0 & 0 & 0 & -\sqrt{\frac{3}{10}} & 0 & 0 \\
 0 & 0 & 0 & -\frac{1}{\sqrt{5}} & -\frac{i}{\sqrt{5}} & 0 & 0 & 0 \\
 0 & 0 & 0 & 0 & 0 & \sqrt{\frac{2}{5}} & 0 & 0 \\
 0 & 0 & 0 & -\frac{1}{\sqrt{5}} & \frac{i}{\sqrt{5}} & 0 & 0 & 0 \\
 \frac{1}{\sqrt{30}} & \frac{i}{\sqrt{30}} & 0 & 0 & 0 & 0 & 0 & 0 \\
 0 & 0 & -\frac{1}{\sqrt{30}} & 0 & 0 & 0 & 0 & \frac{3}{\sqrt{10}} \\
 0 & 0 & -\frac{1}{\sqrt{30}} & 0 & 0 & 0 & 0 & \frac{3}{\sqrt{10}} \\
 \frac{1}{\sqrt{30}} & -\frac{i}{\sqrt{30}} & 0 & 0 & 0 & 0 & 0 & 0 \\
 0 & 0 & 0 & 0 & 0 & 0 & 0 & 0 \\
 0 & 0 & 0 & 0 & 0 & 0 & 0 & 0 \\
 0 & 0 & 0 & 0 & 0 & 0 & 0 & 0 \\
 0 & 0 & 0 & 0 & 0 & 0 & 0 & 0 \\
 0 & 0 & 0 & 0 & 0 & 0 & 0 & 0 \\
 0 & 0 & 0 & 0 & 0 & 0 & 0 & 0 \\
 -\frac{1}{\sqrt{6}} & -\frac{i}{\sqrt{6}} & 0 & 0 & 0 & 0 & 0 & 0 \\
 0 & 0 & \sqrt{\frac{2}{3}} & 0 & 0 & 0 & 0 & 0 \\
 \frac{1}{\sqrt{2}} & -\frac{i}{\sqrt{2}} & 0 & 0 & 0 & 0 & 0 & 0 \\
 \frac{1}{\sqrt{2}} & \frac{i}{\sqrt{2}} & 0 & 0 & 0 & 0 & 0 & 0 \\
 0 & 0 & \sqrt{\frac{2}{3}} & 0 & 0 & 0 & 0 & 0 \\
 -\frac{1}{\sqrt{6}} & \frac{i}{\sqrt{6}} & 0 & 0 & 0 & 0 & 0 & 0 \\
 0 & 0 & 0 & 0 & 0 & 0 & 0 & 0 \\
 0 & 0 & 0 & 0 & 0 & 0 & 0 & 0 \\
 0 & 0 & 0 & \frac{1}{\sqrt{2}} & \frac{i}{\sqrt{2}} & 0 & 0 & 0 \\
 0 & 0 & 0 & 0 & 0 & 1 & i & 0 \\
 0 & 0 & 0 & 0 & 0 & 1 & -i & 0 \\
 0 & 0 & 0 & \frac{1}{\sqrt{2}} & -\frac{i}{\sqrt{2}} & 0 & 0 & 0 \\
 0 & 0 & 0 & 0 & 0 & 0 & 0 & 0
\end{array}
\right) \, ,
\\ \nonumber {\bf T}_{8\times27}^{7\dagger} &=& \left(
\begin{array}{cccccccc}
 0 & 0 & 0 & 0 & 0 & 0 & -\sqrt{\frac{3}{10}} & 0 \\
 0 & 0 & 0 & \frac{i}{\sqrt{5}} & -\frac{1}{\sqrt{5}} & 0 & 0 & 0 \\
 0 & 0 & 0 & 0 & 0 & 0 & \sqrt{\frac{2}{5}} & 0 \\
 0 & 0 & 0 & -\frac{i}{\sqrt{5}} & -\frac{1}{\sqrt{5}} & 0 & 0 & 0 \\
 \frac{i}{\sqrt{30}} & -\frac{1}{\sqrt{30}} & 0 & 0 & 0 & 0 & 0 & 0 \\
 0 & 0 & -\frac{i}{\sqrt{30}} & 0 & 0 & 0 & 0 & \frac{3 i}{\sqrt{10}} \\
 0 & 0 & \frac{i}{\sqrt{30}} & 0 & 0 & 0 & 0 & -\frac{3 i}{\sqrt{10}} \\
 -\frac{i}{\sqrt{30}} & -\frac{1}{\sqrt{30}} & 0 & 0 & 0 & 0 & 0 & 0 \\
 0 & 0 & 0 & 0 & 0 & 0 & 0 & 0 \\
 0 & 0 & 0 & 0 & 0 & 0 & 0 & 0 \\
 0 & 0 & 0 & 0 & 0 & 0 & 0 & 0 \\
 0 & 0 & 0 & 0 & 0 & 0 & 0 & 0 \\
 0 & 0 & 0 & 0 & 0 & 0 & 0 & 0 \\
 0 & 0 & 0 & 0 & 0 & 0 & 0 & 0 \\
 -\frac{i}{\sqrt{6}} & \frac{1}{\sqrt{6}} & 0 & 0 & 0 & 0 & 0 & 0 \\
 0 & 0 & i \sqrt{\frac{2}{3}} & 0 & 0 & 0 & 0 & 0 \\
 \frac{i}{\sqrt{2}} & \frac{1}{\sqrt{2}} & 0 & 0 & 0 & 0 & 0 & 0 \\
 -\frac{i}{\sqrt{2}} & \frac{1}{\sqrt{2}} & 0 & 0 & 0 & 0 & 0 & 0 \\
 0 & 0 & -i \sqrt{\frac{2}{3}} & 0 & 0 & 0 & 0 & 0 \\
 \frac{i}{\sqrt{6}} & \frac{1}{\sqrt{6}} & 0 & 0 & 0 & 0 & 0 & 0 \\
 0 & 0 & 0 & 0 & 0 & 0 & 0 & 0 \\
 0 & 0 & 0 & 0 & 0 & 0 & 0 & 0 \\
 0 & 0 & 0 & \frac{i}{\sqrt{2}} & -\frac{1}{\sqrt{2}} & 0 & 0 & 0 \\
 0 & 0 & 0 & 0 & 0 & i & -1 & 0 \\
 0 & 0 & 0 & 0 & 0 & -i & -1 & 0 \\
 0 & 0 & 0 & -\frac{i}{\sqrt{2}} & -\frac{1}{\sqrt{2}} & 0 & 0 & 0 \\
 0 & 0 & 0 & 0 & 0 & 0 & 0 & 0
\end{array}
\right) \, ,
\\ \nonumber {\bf T}_{8\times27}^{8\dagger} &=& \left(
\begin{array}{cccccccc}
 0 & 0 & 0 & 0 & 0 & 0 & 0 & 3 \sqrt{\frac{3}{10}} \\
 \sqrt{\frac{3}{5}} & i \sqrt{\frac{3}{5}} & 0 & 0 & 0 & 0 & 0 & 0 \\
 0 & 0 & \sqrt{\frac{6}{5}} & 0 & 0 & 0 & 0 & 0 \\
 \sqrt{\frac{3}{5}} & -i \sqrt{\frac{3}{5}} & 0 & 0 & 0 & 0 & 0 & 0 \\
 0 & 0 & 0 & \frac{3}{\sqrt{10}} & \frac{3 i}{\sqrt{10}} & 0 & 0 & 0 \\
 0 & 0 & 0 & 0 & 0 & \frac{3}{\sqrt{10}} & \frac{3 i}{\sqrt{10}} & 0 \\
 0 & 0 & 0 & 0 & 0 & \frac{3}{\sqrt{10}} & -\frac{3 i}{\sqrt{10}} & 0 \\
 0 & 0 & 0 & \frac{3}{\sqrt{10}} & -\frac{3 i}{\sqrt{10}} & 0 & 0 & 0 \\
 0 & 0 & 0 & 0 & 0 & 0 & 0 & 0 \\
 0 & 0 & 0 & 0 & 0 & 0 & 0 & 0 \\
 0 & 0 & 0 & 0 & 0 & 0 & 0 & 0 \\
 0 & 0 & 0 & 0 & 0 & 0 & 0 & 0 \\
 0 & 0 & 0 & 0 & 0 & 0 & 0 & 0 \\
 0 & 0 & 0 & 0 & 0 & 0 & 0 & 0 \\
 0 & 0 & 0 & 0 & 0 & 0 & 0 & 0 \\
 0 & 0 & 0 & 0 & 0 & 0 & 0 & 0 \\
 0 & 0 & 0 & 0 & 0 & 0 & 0 & 0 \\
 0 & 0 & 0 & 0 & 0 & 0 & 0 & 0 \\
 0 & 0 & 0 & 0 & 0 & 0 & 0 & 0 \\
 0 & 0 & 0 & 0 & 0 & 0 & 0 & 0 \\
 0 & 0 & 0 & 0 & 0 & 0 & 0 & 0 \\
 0 & 0 & 0 & 0 & 0 & 0 & 0 & 0 \\
 0 & 0 & 0 & 0 & 0 & 0 & 0 & 0 \\
 0 & 0 & 0 & 0 & 0 & 0 & 0 & 0 \\
 0 & 0 & 0 & 0 & 0 & 0 & 0 & 0 \\
 0 & 0 & 0 & 0 & 0 & 0 & 0 & 0 \\
 0 & 0 & 0 & 0 & 0 & 0 & 0 & 0
\end{array}
\right) \, .
\end{eqnarray}
%
%

\end{document}